\documentclass[prd,amssymb,amsmath,amsfonts,superscriptaddress,nofootinbib,reprint,showpacs,floatfix]{revtex4-1}

\usepackage{graphicx}
\usepackage{lmodern}
\usepackage{amsmath,amssymb}
\usepackage{mathrsfs}
\usepackage{amsfonts}
\usepackage[utf8]{inputenc}
\usepackage{url}
\usepackage[colorlinks]{hyperref}
\usepackage{multirow}
\usepackage{cancel}
\usepackage[x11names,svgnames,rgb,table]{xcolor}
\usepackage[nolist,nohyperlinks]{acronym}
\usepackage{xspace}
\usepackage{float}
\usepackage{makecell,tabularx,colortbl}
\usepackage{booktabs,cellspace}
\usepackage{ulem}
\newcommand{\Hs}{\mathcal{S}}

\newcommand{\Bcal}{\mathcal{B}}
\newcommand{\Hnots}{\cancel{\mathcal{S}}}

\newcommand{\Hg}{\mathcal{G}}
\newcommand{\Hnotg}{\cancel{\Hg}}
\newcommand{\Hn}{\mathcal{N}}

\newcommand{\oddsratio}{\mathcal{O}}
\newcommand{\dataall}{\mathbf{d}}

\newcommand{\pastro}{P(\Hs | \dataall)}
\newcommand{\omicron}{\textsc{Omicron}\xspace}

\newcolumntype{C}{>{\centering\arraybackslash}X}
\newcolumntype{L}{>{\arraybackslash}X}

\definecolor{dodgerblue}{HTML}{1E90FF}
\definecolor{viennared}{HTML}{DA0A14}
\definecolor{ctorange}{HTML}{FF6C0C}
\definecolor{wales}{HTML}{ff0038}
\definecolor{benettongreen}{HTML}{009421}
\definecolor{valenciacfred}{HTML}{ee3524}
\definecolor{barcelonafcgold}{HTML}{edbb00}
\definecolor{jam}{HTML}{A50B5E}
\definecolor{austriawien}{HTML}{441678}
\definecolor{italia90green}{HTML}{009966}
\definecolor{ferrarired}{HTML}{ff2800}
\definecolor{gray}{HTML}{F0F0F0}
\definecolor{LightCyan}{rgb}{0.88,1,1}
\newcolumntype{a}{>{\columncolor{gray}}c}
\newcolumntype{b}{>{\columncolor{white}}c}

\AtBeginDocument{
  \hypersetup{
    citecolor=dodgerblue,
    linkcolor=dodgerblue,   
    urlcolor=dodgerblue
    }
}

\hypersetup{
     colorlinks=true,
     linkcolor=dodgerblue,
     filecolor=dodgerblue,
     citecolor = dodgerblue,      
     urlcolor=dodgerblue,
     }

\newcommand{\Birmingham}{School of Physics and Astronomy and Institute for Gravitational Wave Astronomy, University of Birmingham, Edgbaston, Birmingham, B15 2TT, United Kingdom}
\begin{document}
%%%%%%%%%%

\title{Assessing gravitational-wave binary black hole candidates with Bayesian odds}

\author{Geraint Pratten}
\email{g.pratten@bham.ac.uk}
\affiliation{\Birmingham}

\author{Alberto Vecchio}
\email{av@star.sr.bham.ac.uk}
\affiliation{\Birmingham}

\date{\today}

%\begin{flushright}
%LIGO-P
%\end{flushright}

%%%%%%%%%%%%%%%
\begin{abstract}
Gravitational waves from the coalescence of binary black holes can be distinguished from noise transients in a detector network through Bayesian model selection by exploiting the coherence of the signal across the network. We present a Bayesian framework for calculating the posterior probability that a signal is of astrophysical origin, agnostic to the specific search strategy, pipeline or search domain with which a candidate is identified. We apply this framework under \textit{identical} assumptions to all events reported in the LIGO-Virgo GWTC-1 catalog, GW190412 and numerous event candidates reported by independent search pipelines by other authors. With the exception of GW170818, we find that all GWTC-1 candidates, and GW190412, have odds overwhelmingly in favour of the astrophysical hypothesis, including GW170729, which was assigned significantly different astrophysical probabilities by the different search pipelines used in GWTC-1. GW170818 is de-facto a single detector trigger, and is therefore of no surprise that it is disfavoured as being produced by an astrophysical source in our framework. We find \textit{three} additional event candidates, GW170121, GW170425 and GW170727, that have significant support for the astrophysical hypothesis, with a probability that the signal is of astrophysical origin of 0.53, 0.74 and 0.64 respectively. We carry out a hierarchical population study which includes these three events in addition to those reported in GWTC-1, finding that the main astrophysical results are unaffected.
\end{abstract}

%\pacs{
%04.80.Nn, % gravitational wave detectors and experiments
%95.85.Sz, % Gravitational waves: astronomical observations
%97.80.-d   % Stars: binary and multiple
%04.30.Db, % GW Wave generation and sources
%04.30.Tv,  % GW Gravitational-wave astrophysics
%97.60.Jd, % Neutron stars
%26.60.Kp  % NS EoS
%}

%%%%%%%%%%%%%%%
\maketitle 

%%%%%%%%%%%%%%% INTRODUCTION
\section{Introduction}
\label{sec:intro}
Gravitational-wave (GW) astronomy is having a profound impact on our understanding of the fundamental nature of astrophysial binary black holes and the properties of the underlying population. During the first (O1) and second (O2) observing runs of Advanced LIGO \cite{TheLIGOScientific:2014jea} and Advanced Virgo \cite{TheVirgo:2014hva}, ten unambiguous binary black hole (BBH) mergers were reported as part of the first Gravitational-wave Transient catalog (GWTC-1). Since then, over 25 additional BBH candidates have been reported~\cite{2-OGC, Nitz:2020naa, Zackay:2019btq,Zackay:2019kkv,Zackay:2019tzo,Venumadhav:2019tad,Venumadhav:2019lyq}. This population is expected to grow considerably starting with the events observed during the third observing run (O3), with two confident detections having already been announced~\cite{LIGOScientific:2020stg, Abbott:2020khf}. 

One of the key goals of GW searches is to determine which of the detection candidates are produced by astrophysical compact binaries. This result is often condensed into a single quantity referred to as $p_{\rm astro}$, the probability that a transient signal is of astrophysical origin \cite{Farr:2015fgmc,Kapadia:2019uut,Gaebel:2018poe}. One of the key limiting factors in our ability to confidently identify such astrophysical binaries is the presence of instrumental noise transients (glitches) that contaminate the data. Noise transients can mimic astrophysical signals, impeding the statistical significance to which we can detect GW signals. Advanced LIGO and Advanced Virgo are instruments of exquisite sensitivity with many of the significant detections being observed with signal-to-noise ratios (SNR), $\rho$, so large that $p_{\rm astro}$ is effectively indistinguishable from unity. This is, for example, the case for GW150914~\cite{Abbott:2016blz, Abbott:2016nhf}, the very first BBH merger observed, as well as all but one (GW170729) of the events reported in GWTC-1. 

Under the assumption that coalescing binary systems are distributed uniformly in comoving volume, the expected distribution of signal-to-noise ratio scales as $\sim \rho^{-4}$ \cite{Schutz:2011tw,Chen:2014yla}, but the distribution of background and transient instrumental signals also rises steeply as one goes to progressively smaller values of $\rho$ to identify event candidates at ``threshold" \cite{Nuttall:2015dqa,TheLIGOScientific:2016zmo,TheLIGOScientific:2017lwt,Cabero:2019orq,Davis:2020nyf}. As modelled search pipelines, currently based on the same underlying technique of matched-filtering, battle to identify quiet signals overwhelmed by noise, different choices are made in different studies: for example cuts bases on data quality, signal consistency tests, power spectral density estimations, and physical search parameter domains (e.g. binary mass range covered by a search). 

The astrophysical probability $p_{\rm astro}$ introduced in \cite{Farr:2015fgmc,Kapadia:2019uut,Gaebel:2018poe}, is a Bayesian odds comparing the astrophysical and terrestrial hypotheses. This method estimates the joint posterior on the Poisson expected counts for an arbitrary choice of foreground categories, such as astrophysical BBHs. A key limitation, however, is that the method appeals to bootstrap techniques in order to characterize the noise model for a specific search pipeline. As a consequence of the different choices made by different searches, a possible outcome, in particular for low signal-to-noise events, is that the same signal can be, and in fact has been, assigned a different value of $p_{\rm astro}$ depending on the pipeline and/or search employed. This is the case for GW170729, which was assigned a $p_{\rm astro}$ of $0.98$ by the GstLAL \cite{Messick:2017gst,Sachdev:2019vvd} pipeline and $0.52$ by the PyCBC pipeline \cite{Usman:2015kfa,Nitz:2017svb} in the analyses for GWTC-1 \cite{LIGOScientific:2018mvr}. Possibly more disturbing is that the region of parameter space on which a search is performed -- the ``BBH template bank" versus the full bank covering the whole neutron star and black hole mass range -- can lead to different values of $p_{\rm astro}$ for the same event, to the extent that some signals cross the (arbitrary) threshold of detection in one search and not the other. Although the reason for this behaviour is understood, it is important to understand whether $p_\mathrm{astro}$ is a sufficiently informative and robust characterisation of the probability that detection candidates identified by a search are of astrophysical origin, and whether or not they are included in further studies, either of the fundamental physics properties of black holes \cite{TheLIGOScientific:2016src,TheLIGOScientific:2016pea,LIGOScientific:2019fpa} or the underlying astrophysical properties of the population \cite{TheLIGOScientific:2016htt,TheLIGOScientific:2016pea,LIGOScientific:2018jsj}.

Here we reconsider the problem of assigning a measure of the likelihood of a signal being of astrophysical origin by considering a different statistical quantity: the posterior odds of a transient candidate being produced by an astrophysical source (specifically a BBH described by general relativity) versus random transient noise in the detectors' data, which could arise from instrumental glitches and/or Gaussian noise fluctuations. From this quantity we can trivially evaluate the posterior probability of a signal being of astrophysical origin. Our fully Bayesian framework provides a single, unified measure of the astrophysical probability that is crucially agnostic to the specific search strategy, pipeline implementation or search domain, with which a candidate is identified. Our approach builds on several studies and preliminary investigations~\cite{VeitchVecchio:2010, Smith:2017vfk, Isi-et-al:2018, Ashton:2018jfp, AshtonThraneSmith:2019} and critically takes advantage of observations carried out with a network of detectors: a signal produced by an astrophysical source must appear in the data of the detectors in the network with consistent astrophysical parameters, arrival times, signal strength, and so on.

Using \textit{identical} assumptions, we analyze BBH events reported in GWTC-1~\cite{LIGOScientific:2018mvr} and those reported by other search pipelines, notably the PyCBC 2-OGC catalog~\cite{2-OGC} and the IAS search~\cite{Venumadhav:2019lyq}. We also analyse GW190412, the first BBH detected during the LIGO-Virgo third observing run~\cite{LIGOScientific:2020stg}. The main results are summarised in Table~\ref{tab:triggers} and Figs.~\ref{fig:PS_vs_PSd}, ~\ref{fig:PG_vs_PSd} and~\ref{fig:alpha_beta}. We find that all significant BBHs reported in GWTC-1, modulo the single detector event GW170818, and GW190412 have odds overwhelmingly in favor of the astrophysical hypothesis. For all the other event candidates, the $p_\mathrm{astro}$ reported by the searches is not a monotonic function of the odds. Three candidate events \cite{Venumadhav:2019lyq}, GW170121, GW170727 and GW170425 have odds that favour the astrophysical origin hypothesis and that are sufficient to be included in typical population studies. Interestingly GW170425, which was originally reported with $p_\mathrm{astro} = 0.21$, has odds comparable to GW170727, which was reported with $p_\mathrm{astro} = 0.99$. In order to gauge the astrophysical implications of these new events, we perform a hierarchical population analysis incorporating these new binaries, finding that the main astrophysical results reported in \cite{LIGOScientific:2018jsj} are unchanged. Whilst we focus on BBH coalescences in this paper, the framework can be trivially extended to an arbitrary choice of foreground categories.  

This paper is organised as follows: In Sec.~\ref{sec:notation} we define the model hypotheses and define the two key quantities in our analysis, the posterior odds ratio $\mathcal{O}_{S/N}$ and the probability that the signal is of astrophysical origin $\pastro$. We then discuss the choice of astrophysical signal and glitch priors followed by details regarding the calculation of the evidences and Bayes factors. Section~\ref{sec:results} presents the core results of the paper and Sec.~\ref{sec:poppe} discusses the implications for population inference. We conclude in Sec.~\ref{sec:discussion}. 

%%%%%%%%%%%%%%% 
\section{Astrophysical signal odds}
\label{sec:notation}
The end-point of our analysis is to evaluate the posterior odds ratio, $\oddsratio_{S/N}$,  that an event candidate is due to an astrophysical source -- a BBH merger -- versus a transient noise fluctuation, and as a consequence the posterior probability, $P(\Hs| \dataall)$,  that an astrophysical signal is present in the data.

Hence, given the data $\dataall$, we wish to compute
\begin{align}
    \oddsratio_{S/N} &= \frac{P(\Hs| \dataall)}{P(\Hn| \dataall)}\,,
    \label{eq:odds0}
\end{align}
where $P(\Hs| \dataall)$ and $P(\Hn| \dataall)$ are the posterior probability of the signal hypothesis, $\Hs$, and noise hypothesis, $\Hn$, respectively. The data $\dataall = \{d_k; k = 1,\dots N_d\}$ comprise the collection of data sets, $d_k$, from $N_d$ detectors in the network.
 
As the signal and noise hypotheses are exhaustive, we can use Eq.~(\ref{eq:odds0}) to determine the probability that the signal is of astrophysical origin
\begin{align}
P(\Hs | \dataall) &= \frac{\mathcal{O}_{S/N}}{1 \; + \; \mathcal{O}_{S/N}}\,.
\label{eq:P_signal}
\end{align}
where Eq.~(\ref{eq:P_signal}) is the quantity returned by our analysis that plays a similar role to the $p_{\rm astro}$ computed from the distribution of background and foreground events produced by the search pipelines~\cite{Farr:2015fgmc,Abbott:2016nhf,Kapadia:2019uut,Zackay:2019btq}. However, $\oddsratio_{S/N}$ and $\pastro$ are completely independent of the search pipeline used to identify a candidate, and only depend on the model hypotheses and the associated prior probabilities. They are derived by naturally using a \textit{coherent} analysis of the data -- as opposed to searches, which rely on identifying triggers in \textit{coincidence} -- and can implement the best signal and noise model at one's disposal.

We note that similar strategies for computing Eq.~(\ref{eq:P_signal}) have been pursued in \cite{Smith:2017vfk,AshtonThraneSmith:2019}. However, the signal odds, as defined here, does not rely on the marginalization over a glitch hyper model using contextual data \cite{Smith:2017vfk,AshtonThraneSmith:2019} nor is it treated as a traditional detection statistic to obtain a frequentist estimate of the significance of an event given the measured background as in~\cite{Isi-et-al:2018}. 

\subsection{Notation and assumptions}
Here we summarise the assumptions for the computation of Eq.~(\ref{eq:odds0}) and~(\ref{eq:P_signal}) and summarise our notation, which closely follows~\cite{VeitchVecchio:2010, Isi-et-al:2018, Ashton:2018jfp,AshtonThraneSmith:2019}, to which we refer the reader for further details.

First, we define the models that we consider:
\begin{itemize}
    \item \textbf{Signal model or hypothesis}, $\Hs$: \textit{There is an astrophysical signal due to a BBH coalescence in the detector network}. We also make the additional assumption that no transient of instrumental nature -- a ``glitch" -- takes place at the same time of the GW signal. This is an excellent approximation in the case of BBHs, the focus of this study, at current instrument sensitivity, as GWs from BBH coalescences are in the instruments' bandwidth for at best a few seconds. The signal model corresponds therefore to $\Hs = \Hs_1 \land \dots \land \Hs_{N_d}$, where $\land$ is the logical ``and".  
    \item \textbf{Noise model or hypothesis}, $\Hn$: \textit{there is no astrophysical signal in the data, just noise}. To an excellent approximation, we can further assume that the noise between two detectors is uncorrelated, see~\cite{LIGOScientific:2019hgc} and references therein. The noise in each of the instruments could be due to a glitch -- hypothesis $\Hg_k$ -- or simply Gaussian stationary noise, a model that we identify with $\Hnotg_k$. Therefore $\Hg_k$ and $\Hnotg_k$ are exhaustive and disjoint noise hypotheses. We also make an important additional assumption: the glitch model ($\Hg_k$) for each individual detector is the most conservative one in which glitches are modelled as having the same functional form as a GW from a BBH coalescence with uncorrelated parameters in each detector, $\Hg_k = \Hs_k$ \cite{VeitchVecchio:2010}. Under this assumption, the noise model in each instrument is $\Hn_k = \Hnots_k \land (\Hg_k \lor \Hnotg_k)$, where $\lor$ is the logical or. The noise hypothesis can then be written as $\Hn = \Hn_1 \land \dots \land \Hn_{N_d}$.
\end{itemize}
We can now express the odds ratio, Eq.~(\ref{eq:odds0}), in terms of the prior probability of the signal $P(\Hs)$ and noise $P(\Hn)$ hypotheses, and of the marginal likelihoods or evidences, $P(\dataall | \Hs)$ and $P(\dataall | \Hn)$, of the same models~\cite{VeitchVecchio:2010, Isi-et-al:2018}:
\begin{align}
    \oddsratio_{S/N} &= \left[ \frac{P(\Hs)}{P(\Hn)} \right] \, \frac{P(\dataall | \Hs)}{P(\dataall | \Hn)},
    \nonumber \\
    &= 
    \left[\frac{P(\Hs)}{P(\Hn)}\right]\,
    \frac{\Bcal{s/n}}{\prod_{k}\left\{ P(\Hg_k | \Hn)\,\Bcal_{s/n}^{(k)} +  \left[1 - P(\Hg_k | \Hn)\right]\right\}}\,.
    \label{eq:odds_PS_PN}
\end{align}
Here we have defined the single-detector Bayes factor as
\begin{align}
    \mathcal{B}_{s/n}^{(k)} \equiv \frac{P(d_k | \Hs_k)}{P(d_k | \Hnotg_k)}\,,
    \label{eq:B_sn_k}
\end{align}
and the Bayes factor for a coherent GW signal across the network embedded in stationary Gaussian noise as
\begin{align}
    \mathcal{B}_{s/n} \equiv \frac{P(\dataall | \Hs)}{P(\dataall | \Hnotg)}\,,
    \label{eq:B_sn}
\end{align}
where $P(\dataall | \Hnotg) = \prod_k  P(d_k | \Hnotg_k)$.

There are a number of limiting cases of Eq.~(\ref{eq:odds_PS_PN}) that have been discussed in the literature and used in GW data analysis. First, if we assume the case of a perfect instrument, where the noise is Gaussian and stationary and no glitches occur, $\Hn = \Hnotg$, hence $P(\Hg_k | \Hn) = 0$. Consequentially, Eq~(\ref{eq:odds_PS_PN}) reduces to
\begin{equation}
    \oddsratio_{S/N} = 
    \left[\frac{P(\Hs)}{P(\Hn)}\right]\,\Bcal{s/n}, 
    \label{eq:bsn}
\end{equation}
which reduces to the signal vs. Gaussian noise Bayes factor in the limit $P(\Hs) = P(\Hn)$. If we now assume a fundamentally flawed detector, in which the glitch rate is sufficiently high that it can be well approximated by $P(\Hg_k | \Hn) = 1$, then Eq~(\ref{eq:odds_PS_PN}) becomes
\begin{equation}
    \oddsratio_{S/N} = \left[\frac{P(\Hs)}{P(\Hn)}\right]\frac{\Bcal{s/n}}{\prod_{k}\Bcal_{s/n}^{(k)}}\,,
\end{equation}
which reduces to the ``coherent vs incoherent" Bayes factor~\cite{VeitchVecchio:2010}
\begin{equation}
    \Bcal_{c/i} \equiv \frac{\Bcal{s/n}}{\prod_{k}\Bcal_{s/n}^{(k)}} ,
    \label{eq:B_CI}
\end{equation}
under the assumption that $P(\Hs) = P(\Hn)$. If we next assume that 
\begin{equation}
    P(\Hg_k | \Hn) = 0.5 \quad \forall k,
\end{equation}
then the odds ratio reduces to 
\begin{equation}
    \oddsratio_{S/N} = \left[\frac{P(\Hs)}{P(\Hn)}\right]\frac{2\Bcal{s/n}}{\prod_{k}\left(\Bcal_{s/n}^{(k)} + 1 \right)}
\end{equation}
which is related to the ``coherent vs incoherent or noise" Bayes' factor
\begin{equation}
    \Bcal_{c/(i\lor n)} \equiv
    \frac{\Bcal{s/n}}{\prod_{k}\left(\Bcal_{s/n}^{(k)} + 1 \right)}
    \label{eq:B_CIorN}\,.
\end{equation} 
Finally, if one sets $\alpha \equiv P(\Hs)/P(\Hn) \simeq P(\Hs) = 10^{-6}$ and $\beta = P(\Hg_k | \Hn) = 10^{-4}$ $(\forall k)$, the odds ratio reduces to the ``Bayesian coherence ratio" introduced in~\cite{Isi-et-al:2018},
\begin{align}
\label{eq:bcr}
    \mathcal{B}_{\rm cr} = \alpha\,\frac{\Bcal{s/n}}{\prod_{k}\left\{\beta\,\Bcal_{s/n}^{(k)} +  \left[1 - \beta\right]\right\}} ,
\end{align}
where the values of $\alpha$ and $\beta$ were chosen through an injection campaign aimed at separating the foreground and background populations.

\subsection{Priors for signal and glitch models}
\label{sec:priors}
An important aspect of our framework is the choice of model priors, which affect the results in a straightforward way, see Eq.~(\ref{eq:odds_PS_PN}). The signal odds scales linearly with the prior belief of an astrophysical signal being present in a given segment of data, ${P(\Hs)}/{P(\Hn)}$, while the glitch probability acts as a weighting factor for the presence of an uncorrelated transient in each of the detectors in the network. 

As discussed in Sec~\ref{sec:analysis}, for each of the event candidates we select an $8$s data segment containing the putative signal. We set a uniform prior on the coalescence time, $t_c$, to search for a BBH merger occuring in an interval $\Delta t_c = 0.2$s around the GPS time of the reported event candidate. As BBH mergers are rare~\cite{LIGOScientific:2018mvr}, we can assume that $P(\Hs) \ll 1$. Leveraging on prior knowledge of the BBH merger rate, we set $P(\Hs)$ to be the probability of a coalescence occurring in an interval $\Delta t_c$ and be produced by a binary within the sensitive spacetime volume for all signals that yield a single interferometer SNR $\rho \geq 7$ for the typical detector sensitivities throughout the observing period. The SNR threshold used here is slightly lower than the more conventional $\rho > 8$ \cite{Finn:1992xs} as we are particularly interested in signals at the detection threshold, whilst avoiding the steep rise in the background of glitches. Using these assumptions, we have  
\begin{equation}
\label{eq:p_S}
        P(\Hs) = 6.43 \times 10^{-7} \left( \frac{\mathcal{R}_0}{53.2 \,\mathrm{Gpc}^{-3}\,\mathrm{y}^{-1}} \right) \left( \frac{\Delta t_c}{0.2\,\mathrm{s}} \right) \,,
\end{equation}
where $\mathcal{R}_0$ is the local merger rate of BBHs; we provide in App.~\ref{app:sig_rate} further details about how this result is derived. 

One may wonder whether using the value of $\mathcal{R}_0$ derived from the analysis of all the data in GWTC-1 is formally consistent with determining $P(\Hs)$ when analysing all other event candidates identified during the same observing period. 
A more formally correct strategy, that we plan to implement in future analyses, would be to to divide the data into segments -- say 1-week long -- analyse the data, and based on the identified candidates and their astrophysical probability determine the BBH merger rate at that point, and hence update $P(\Hs)$ for the next segment and so forth. In practice, however, the results presented here would be unaffected.
The estimate of the BBH merger rate determined using the initial 16 days surrounding GW150914 -- which would be a natural starting point for any iterative analysis -- is consistent with $\mathcal{R}_0$ measured at the end of O2 within a factor $\approx 2$, though the uncertainty is significantly reduced. Our results, as described in Sec~\ref{sec:results}, are robust against such variations in $P(\mathcal{S})$, see in particular Fig~\ref{fig:PS_vs_PSd}.
 
As the models $\Hs$ and $\Hn$ are exhaustive hypotheses, $P(\Hs) + P (\Hn) = 1$, the prior ratio in Eq.~(\ref{eq:odds_PS_PN}) can be written as
\begin{equation}
        \frac{P(\Hs)}{P(\Hn)} = \frac{P(\Hs)}{1-P(\Hs)} \simeq P(\Hs), 
\end{equation}
to a very good approximation. The effect of $P(\Hs)$ on $\mathcal{O}_{S/N}$ is to act as an overall normalization, with the odds, and hence the posterior astrophysical signal probability, scaling linearly with the total astrophysical merger rate integrated over the entire binary population, as in Eq.~(\ref{eq:p_S}).

In order to compute Eq.~(\ref{eq:odds_PS_PN}), we also need an estimate of the probability that a glitch takes place within the same prior coalescence interval, $P(\Hg_k | \Hn)$. This is a quantity whose value changes during the course of a run and can differ between each of the detectors in the network. Here we determine the glitch prior by considering all glitches identified by the \omicron pipeline \cite{Chatterji:2004qg,robinet:2015omi,Robinet:2020om}. For each detector, we take a 24 hour period centered around the time of the putative candidate and estimate the number of triggers, $N^{(\textsc{Om})}_k$ occurring within the frequency range covered by our analysis, $20 - 1024\,\mathrm{Hz}$, and which also yield a signal-to-noise ratio $\ge 7$, consistent with our choice for the signal prior. We set the prior glitch probability for each instrument to
\begin{align}
    P(\Hg_k | \Hn) = N^{(\textsc{Om})}_k \frac{\Delta t_c}{T^{(\mathrm{live})}_k}\,,
    \label{eq:PGk}
\end{align}
where $T^{(\mathrm{live})}_k$ is the live-time for instrument $k$ during the 24-hour period. Note that transients producing a given signal-to-noise ratio when analysed with the \omicron pipeline would in general yield a lower signal-to-noise ratio when using a BBH waveform; our approach is therefore conservative with respect to the chance that the glitches mimic the GW signal from an astrophysical BBH. The values of $P(\Hg_k | \Hn)$ adopted in our analysis are reported in Table \ref{tab:PGk} and are consistent with previous studies of the glitch rates, \textit{e.g.}~\cite{Nuttall:2015dqa,TheLIGOScientific:2016zmo,TheLIGOScientific:2017lwt,Davis:2020nyf}. Our choice of the signal $P(\Hs)$ and glitch prior $P(\Hg_k | \Hn)$ is also broadly consistent with values for related quantities considered in other studies~\cite{Isi-et-al:2018, Ashton:2020odd}.

\subsection{Evidence and Bayes factor evaluation}
\label{sec:analysis}
The last quantities we need to evaluate Eqs.~(\ref{eq:P_signal}) and ~(\ref{eq:odds_PS_PN}) are the signal evidences, and therefore the Bayes factors. For this, we follow the approach, and use the same software, adopted for the analyses of the BBHs reported in GWTC-1 \cite{LIGOScientific:2018mvr}.

For each event, we use data from the Gravitational Wave Open Science Center \cite{Abbott:2019ebz,GWOSC} and analyze $8$s of data around the time of the candidate and within a frequency range of $20-1024$Hz. For simplicity, we restrict our analysis to a 2 detector network, using only the data from Hanford and Livingston, the two most sensitive instruments\footnote{The BBH candidates for which Virgo was also in science mode are: GW170729, GW170729A, GW170809, GW170814, GW170801, GW170817A and GW170818. Our analysis is completely general and can be trivially (though rather costly in terms of computer time) extended to a network consisting of an arbitrary number of instruments. We leave this to future work, noting that during the first two observing runs Virgo was sufficiently less sensitive than LIGO that the main conclusions presented here are unaffected.}. The power spectral densities (PSDs) are estimated using the \texttt{BayesWave} algorithm \cite{Littenberg:2014oda,Cornish:2014kda} and we marginalize over calibration uncertainty \cite{Vitale:2011wu,Cahillane:2017vkb,Sun:2020wke} using the approximate uncertainty reported in GWTC-1 \cite{LIGOScientific:2018mvr}. Whilst do not include marginalization over PSD uncertainty when performing parameter estimation \cite{Chatziioannou:2019zvs,Banagiri:2019lon,Biscoveanu:2020kat,Talbot:2020auc}, the \texttt{BayesWave} algorithm does marginalize over uncertainty when generating the PSD.

To compute the evidences, we perform a coherent Bayesian analysis using the nested sampling algorithm \cite{Skilling:2006gxv,VeitchVecchio:2008prd,VeitchVecchio:2008prd,VeitchVecchio:2010} implemented in \texttt{LALInference} \cite{Veitch:2014wba}. We use the precessing waveform model IMRPhenomPv2 \cite{Hannam:2013oca,Schmidt:2014iyl,Husa:2015iqa,Khan:2015jqa} for both the signal and incoherent glitch model. 

In order to analyze each event on an equivalent footing, we adopt the same priors for the BBH parameters for all events. The choice of priors used in our analysis is templated on the default settings used in GWTC-1 \cite{LIGOScientific:2018mvr}. We use a uniform prior in the component masses $m_{1,2}$ in the range $[5.0,160.0]\,M_\odot$ and an isotropic spin prior with dimensionless spin magnitudes $\chi_{1,2}$ taken to be within $[0,0.99]$. We further restrict the redshifted chirp mass $(1+z) \mathcal{M} = (m_1 m_2)^{3/5} / (m_1 + m_2)^{1/5}$ to lie within $[5.0,100.0]\,M_\odot$ and the mass-ratio $q = m_2 / m_1$ to lie within $0.05 \leq q \leq 1$. The distance prior is taken to be proportional to the luminosity distance squared, with an upper limit of $5 \, \rm{Gpc}$.

%%%%%%%%%%%%%%% 

\begin{table*} 
\centering

\caption{Table summarising the astrophysical probability $P(\Hs | \dataall)$ and the Bayes factors computed according to different assumptions: $\mathcal{O}_{S/N}$ in Eq.~(\ref{eq:odds_PS_PN}), $\mathcal{B}_{\rm cr}$ in Eq.~(\ref{eq:bcr}) using $\lbrace \alpha = 10^{-6}, \beta_k = 10^{-4} \rbrace$, $\mathcal{B}_{s/n}$ defined in Eq.~(\ref{eq:bsn}), $\mathcal{B}_{c/i}$ in Eq.~(\ref{eq:B_CI}) and $\mathcal{B}_{c/(i.\lor n.)}$ in Eq.~(\ref{eq:B_CIorN}). For $\mathcal{O}_{S/N}$ we use $P(\Hs) = 6.43 \times 10^{-7}$ and $P(\Hg_k | \dataall)$ as estimated from the omciron triggers in a $24$h window around each event and reported in Table~\ref{tab:PGk}. For GW190412 \cite{LIGOScientific:2020stg} we adopt $P(\Hg_k | \Hn) = 10^{-4}$ as a fiducial value, though this event is unambiguous with $\pastro \sim 1$ irrespective of the glitch prior, assuming reasonable values. The purple shaded rows denote the significant events reported in GWTC-1 \cite{LIGOScientific:2018mvr} and the blue shaded rows denote the new event candidates with $\pastro > 0.5$. The $p_{\rm astro}$ values are taken from the following search pipelines: PyCBC ${}^{\ddagger}$ \cite{LIGOScientific:2018mvr}, GstLAL ${}^{\|}$ \cite{LIGOScientific:2018mvr}, ${}^{\dagger}$ \cite{2-OGC}, ${}^{\S}$ \cite{Nitz:2020naa} and ${}^{\ast}$ \cite{Venumadhav:2019lyq,Zackay:2019btq}.
\\
}
\label{tab:triggers}

\begin{tabularx}{0.9\linewidth}{SC@{} SC@{} SC@{} SC@{} SC@{} SC@{} SC@{} SC@{} SC@{}} 
Event & $\rho_{\rm MF}^N$ & $p_{\rm{astro}}$ & $P(\Hs | \dataall)$  & $\log_{10} \mathcal{O}_{s/n}$ & $\log_{10} \mathcal{B}_{cr}$ & $\log_{10} \mathcal{B}_{s/n}$  &  $\log_{10} \mathcal{B}_{c/i}$ &  $\log_{10} \mathcal{B}_{c/(i \vee n)}$ \\[0.07cm]
\hline
\hline
\rowcolor{austriawien!05}
GW150914 &              25.01 &              $0.99^{\|}, 1.00^{\ddagger}$ &                   1.00 &                           7.86 &                          8.87 &                         121.10 &                           6.87 &                                    6.87 \\
\rowcolor{austriawien!05}
GW151012 &               9.63 &              $0.97^{\|}, 0.96^{\ddagger}$ &                   0.99 &                           2.02 &                          2.24 &                           8.25 &                           5.51 &                                    5.51 \\
\rowcolor{austriawien!05}
GW151226 &              12.71 &              $0.88^{\|},1.00^{\ddagger}$ &                   1.00 &                           6.02 &                          7.19 &                          18.45 &                           8.96 &                                    8.96 \\
\rowcolor{austriawien!05}
GW170104 &              14.01 &              $1.00^{\|},1.00^{\ddagger}$ &                   1.00 &                           6.75 &                          8.38 &                          30.19 &                           6.38 &                                    6.38 \\
\rowcolor{austriawien!05}
GW170608 &              15.62 &              $0.92^{\|},1.00^{\ddagger}$ &                   1.00 &                           8.32 &                          9.79 &                          34.70 &                           7.79 &                                    7.79 \\
\rowcolor{austriawien!05}
GW170729 &              10.62 &              $0.98^{\|},0.52^{\ddagger}$ &                   1.00 &                           3.46 &                          4.65 &                          15.52 &                           2.66 &                                    2.66 \\
\rowcolor{austriawien!05}
GW170809 &              12.82 &              $0.99^{\|},1.00^{\ddagger}$ &                   1.00 &                           3.77 &                          4.45 &                          23.44 &                           3.09 &                                    3.09 \\
\rowcolor{austriawien!05}
GW170814 &              16.66 &              $1.00^{\|},1.00^{\ddagger}$ &                   1.00 &                           6.86 &                          7.80 &                          46.32 &                           5.80 &                                    5.80 \\
\rowcolor{austriawien!05}
GW170818 &              11.34 &              $0.99^{\|}$ &                   0.10 &                          -0.94 &                         -0.39 &                          15.58 &                           1.56 &                                    1.56 \\
\rowcolor{austriawien!05}
GW170823 &              12.04 &              $0.99^{\|},1.00^{\ddagger}$ &                   1.00 &                           5.30 &                          5.89 &                          21.72 &                           4.21 &                                    4.21 \\
\hline
GW151011  &               7.46 &             $0.08^{\dagger}$ &                   0.01 &                          -1.93 &                         -1.71 &                           4.30 &                           1.56 &                                    1.56 \\
GW151124  &               8.67 &             $0.16^{\S}$ &                   0.00 &                          -5.69 &                         -4.66 &                           5.10 &                          -2.38 &                                   -2.38 \\
GW151205  &               6.81 &             $0.53^{\dagger}$ &                   0.03 &                          -1.53 &                         -1.33 &                           4.67 &                           2.06 &                                    2.05 \\
GW151216  &               8.16 &             $0.18^{\dagger}$ &                   0.00 &                          -2.35 &                         -2.16 &                           3.84 &                           2.96 &                                    2.91 \\
GW151216A &               8.48 &             $0.07^{\dagger}$ &                   0.00 &                          -6.02 &                         -5.82 &                           0.18 &                           0.34 &                                   -0.05 \\
GW151217  &               7.99 &             $0.26^{\dagger}$ &                   0.00 &                          -6.04 &                         -5.85 &                           0.15 &                           0.22 &                                   -0.12 \\
GW151222  &              10.22 &             $0.03^{\S}$ &                   0.00 &                          -4.68 &                         -3.52 &                           9.34 &                          -1.65 &                                   -1.65 \\
GW170104A &               7.36 &             $0.12^{\dagger}$ &                   0.00 &                          -4.82 &                         -4.63 &                           1.37 &                           0.97 &                                    0.83 \\
GW170106  &               9.62 &             $0.01^{\S}$ &                   0.00 &                          -6.35 &                         -4.96 &                           8.19 &                          -2.88 &                                   -2.88 \\
\rowcolor{dodgerblue!10}
GW170121  &              10.55 &             $1.00^{\dagger},1.00^{\ast}$ &                   0.53 &                           0.05 &                          1.78 &                          12.26 &                           3.38 &                                    3.38 \\
GW170123  &               6.20 &             $0.08^{\dagger}$ &                   0.00 &                          -3.86 &                         -3.66 &                           2.34 &                           1.70 &                                    1.61 \\
GW170201  &               8.25 &             $0.24^{\dagger}$ &                   0.00 &                          -2.55 &                         -2.36 &                           3.64 &                           2.84 &                                    2.77 \\
GW170202  &               8.50 &             $0.13^{\dagger},0.68^{\ast}$ &                   0.12 &                          -0.86 &                         -0.67 &                           5.33 &                           3.54 &                                    3.54 \\
GW170220  &               6.59 &            $0.10^{\dagger}$ &                   0.00 &                          -4.01 &                         -3.82 &                           2.18 &                           1.06 &                                    1.03 \\
GW170304  &               8.50 &             $0.70^{\dagger},0.99^{\ast}$ &                   0.03 &                          -1.52 &                         -0.31 &                           7.53 &                           1.24 &                                    1.24 \\
GW170402  &               9.00 &             $0.68^{\ast}$ &                   0.00 &                          -3.55 &                         -2.52 &                           6.16 &                          -0.53 &                                   -0.53 \\
GW170403  &               8.20 &             $0.03^{\dagger},0.56^{\ast}$ &                   0.27 &                          -0.42 &                         -0.21 &                           5.80 &                           2.45 &                                    2.45 \\
\rowcolor{dodgerblue!10}
GW170425  &               7.99 &             $0.21^{\dagger},0.77^{\ast}$ &                   0.74 &                           0.46 &                          0.72 &                           6.74 &                           2.44 &                                    2.43 \\
GW170620  &               8.36 &             $0.02^{\dagger}$ &                   0.00 &                          -2.81 &                         -2.62 &                           3.38 &                           2.94 &                                    2.80 \\
GW170629  &               7.56 &             $0.02^{\dagger}$ &                   0.00 &                          -6.25 &                         -6.06 &                          -0.06 &                           0.36 &                                   -0.20 \\
GW170721  &               8.55 &             $0.06$ &                   0.15 &                          -0.75 &                         -0.39 &                           5.66 &                           2.35 &                                    2.35 \\
GW170724  &              10.76 &             $0.02^{\S}$ &                   0.00 &                          -6.77 &                         -5.68 &                          13.46 &                          -3.48 &                                   -3.48 \\
\rowcolor{dodgerblue!10}
GW170727  &               9.90 &             $0.99^{\dagger},0.98^{\ast}$ &                   0.66 &                           0.28 &                          0.87 &                           9.15 &                           2.71 &                                    2.71 \\
GW170729A &              11.55 &             $0.05^{\S}$ &                   0.00 &                          -4.95 &                         -4.24 &                          10.50 &                          -2.93 &                                   -2.93 \\
GW170801  &               8.14 &            $-$ &                   0.00 &                          -4.13 &                         -3.70 &                           2.69 &                          -1.53 &                                   -1.53 \\
GW170817A &              10.65 &             $0.86^{\ast}$ &                   0.02 &                          -1.66 &                         -1.26 &                          16.62 &                           0.63 &                                    0.63 \\
\hline
\rowcolor{austriawien!05}
GW190412  &              18.47 &            $-$ &                   1.00 &                           7.94 &                          8.14 &                          59.27 &                           6.14 &                                    6.14 \\
\end{tabularx}
\end{table*}

%%%%%%%%%%%%%%%%%%%%%%%%%%%%%%%%%%%%%%%%%%%

\section{Results}
\label{sec:results}

We analyze the event candidates reported in GWTC-1 \cite{LIGOScientific:2018mvr}, the PyCBC 2-OGC catalog \cite{2-OGC}, the PyCBC single detector search \cite{Nitz:2020naa}\footnote{Note that we only consider triggers that are found in more than one detector.}, and the independent IAS search \cite{Zackay:2019btq,Zackay:2019kkv,Zackay:2019tzo,Venumadhav:2019tad,Venumadhav:2019lyq}. We also analyse GW190412 \cite{LIGOScientific:2020stg}, the first BBH reported from the third observing run. Here, we focus on BBH events, restricting our analysis to event candidates with a redshifted chirp mass $(1+z) \mathcal{M} > 5 M_{\odot}$. In addition, in order to gauge the robustness of our Bayesian framework, we analyse a set of known background glitches identified by the PyCBC \cite{Usman:2015kfa,Nitz:2017svb} and GstLAL \cite{Messick:2017gst,Sachdev:2019vvd} search pipeline. 

In Table~\ref{tab:triggers}, we summarise the various Bayesian measures of significance for all event candidates considered in our analysis. We report the astrophysical signal odds $\mathcal{O}_{S/N}$, defined in Eq.~(\ref{eq:odds_PS_PN}), and the astrophysical signal probability $\pastro$, Eq.~(\ref{eq:P_signal}). The prior for the signal hypothesis is given by Eq.~(\ref{eq:p_S}) and the glitch prior per event for each detector is given in Table~\ref{tab:PGk} of Appendix~\ref{app:glitch_prob}. For ease of comparison with the literature, we also show the value of $p_\mathrm{astro}$ reported by the various search pipelines. In addition we report the Bayes factors under various limiting cases, defined in Eqns.~(\ref{eq:odds_PS_PN}), (\ref{eq:bsn}), (\ref{eq:B_CI}) and (\ref{eq:B_CIorN}). As a further (obvious) confirmation that $p_\mathrm{astro}$ is pipeline specific, it is also useful to stress that ranking candidates based on $\pastro$ produces a list which differs from the equivalent lists based on $p_\mathrm{astro}$.

\subsection{GWTC-1}
For all binaries reported in GWTC-1, bar GW170818 which we discuss below, and GW190412, we find overwhelming evidence in favour of the astrophysical signal hypothesis, $\pastro \sim 1$. For GW170729, which was assigned a $p_\mathrm{astro}$ of 0.52 and 0.98 by the two main search pipelines in \cite{LIGOScientific:2018mvr}, we find an unambiguous astrophysical signal probability of $\pastro = 1$. 

For GW170818 we find $\pastro = 0.1$. This signal was initially detected as a triple-coincidence event by GstLAL with a signal-to-noise ratio of $\approx 10$ in Livingston, and $\approx 4$ in both Hanford and Virgo \cite{LIGOScientific:2018mvr}. We find consistent values of signal-to-noise ratio in our analysis, see Table~\ref{tab:PGk}, though we do not analyse data from Virgo. The event was initially only detected in the Livingston data by the PyCBC search, though a re-analysis using modified settings around the time of the event did find triggers with a similar signal-to-noise ratio to those reported by GstLAL \cite{LIGOScientific:2018mvr}. This event has also been reported in the dedicated BBH search in \cite{2-OGC} and as a significant single-detector trigger in \cite{Venumadhav:2019lyq,Zackay:2019btq}.

It is clear that GW170818 is a relatively weak signal and, due to the markedly different sensitivity of the instruments, is essentially a ``single detector" event. It is therefore unsurprising that the astrophysical signal probability returned by our approach -- which critically relies on the notion of coherent signals across a network of detectors -- is small, and in tension with results reported by the search pipelines. This is expected to be the case whenever the signal in one of the two detectors is too weak to be distinguishable from Gaussian noise ($\rho \sim 5.5$). As further insight into the nature of this candidate, adopting the Bayesian coherence ratio~\cite{Isi-et-al:2018} as a measure of significance, we find $\log_{10} \mathcal{B}_{cr} = -0.39$, the only circumstance in which one obtains a negative value for the significant BBH events in GWTC-1. 

Finally, we note that for single detector triggers, alternative strategies for determining the significance and astrophysical signal odds should be pursued \cite{Callister:2017urp,LIGOScientific:2018mvr,Zackay:2019btq,Nitz:2020naa}. In particular see \cite{Nitz:2020naa}, which reports a large positive Bayes factor for a coherent signal versus a null hypothesis consisting of a signal in one detector and noise in all others.

%%%% alpha-beta curves %%%%
\begin{figure*}[ht!]
    \centering
    \includegraphics[width=\linewidth]{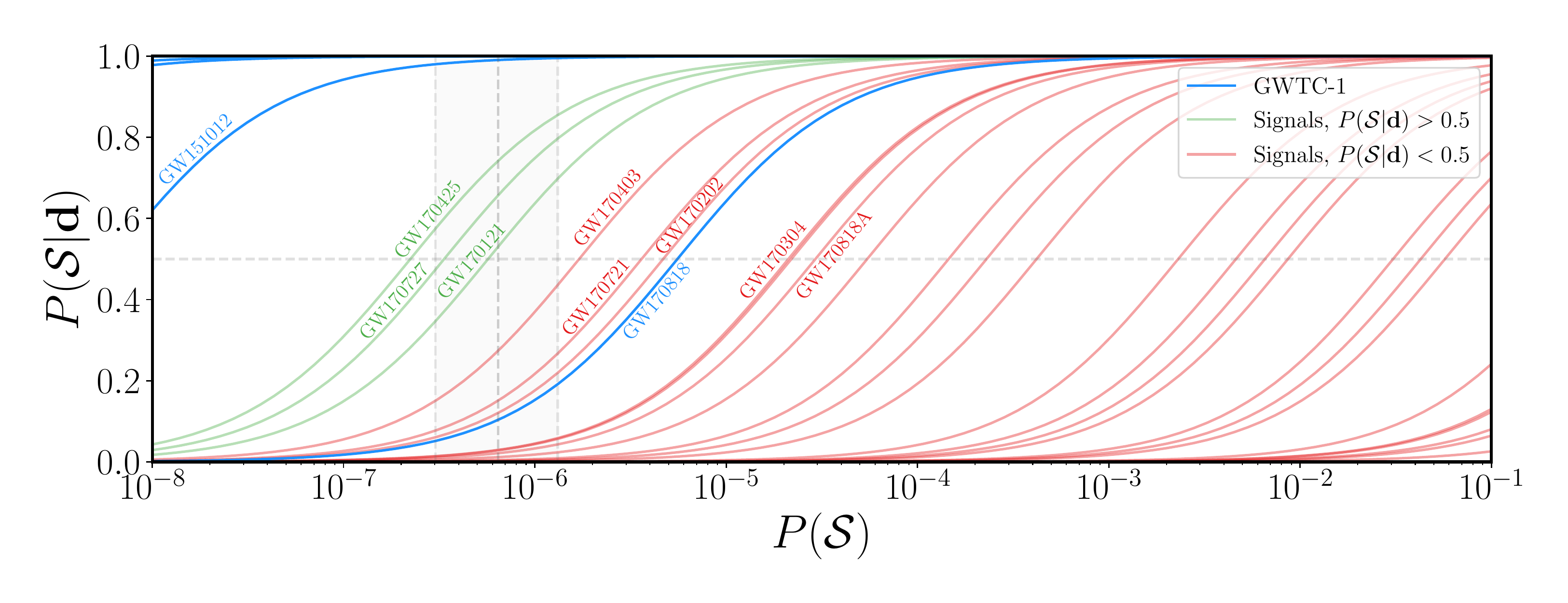}
        \caption{$\pastro$ as a function of $P(\Hs)$. The probability of the glitch hypothesis $P(\Hg_k | \Hn)$ for each detector is estimated using the trigger rate in $24$h of data around each event. The horizontal dashed line denotes the threshold for inclusion in the population analysis, $P(\Hs | \dataall) > 0.5$. The vertical dashed lines denote the median and $90\%$ CI for the signal hypothesis $P(\Hs)$ as inferred from the uncertainty in the local astrophysical merger rate $\mathcal{R}_0$.
        }
    \label{fig:PS_vs_PSd}
\end{figure*}

\begin{figure*}[ht!]
    \centering
    \includegraphics[width=\linewidth]{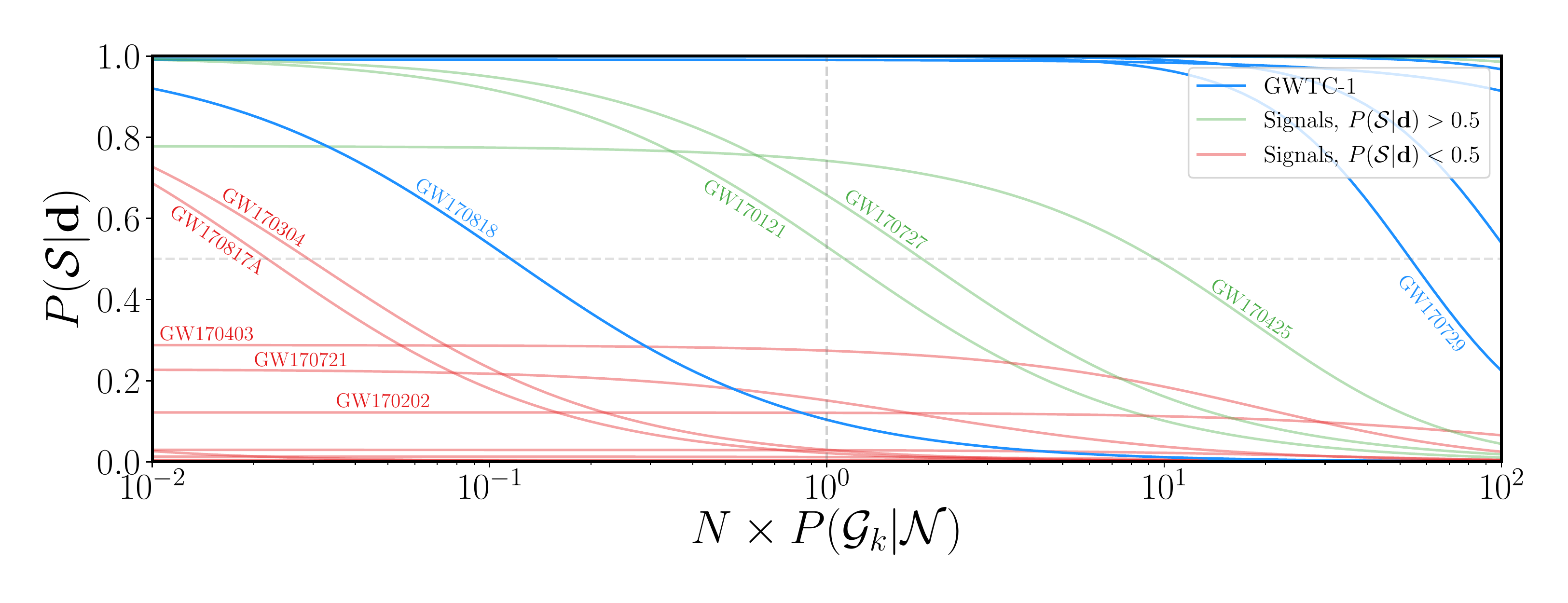}
        \caption{$\pastro$ as a function of the glitch probability $P(\Hg_k | \Hn)$. Here we scale the fiducial values $P(\Hg_k | \Hn)$ by a factor $N \in \left[ 10^{-2}, 10^{2} \right]$, keeping the ratio of the glitch probability in each detector fixed, i.e. $P(\Hg_H | \Hn) / P(\Hg_L | \Hn) = \rm{const}$. For all GWTC-1 events (blue), bar the single detector event GW170818, we would need to increase the glitch probability by a factor $\sim \mathcal{O}(10^2)$ in order for the astrophysical probability $\pastro$ to drop below $0.5$. Of the interesting event candidates (green), GW170121 is particularly sensitive to the glitch probability, with even a small increase in $P(\Hg_k | \Hn)$ pushing $\pastro$ below $0.5$.}
    \label{fig:PG_vs_PSd}
\end{figure*}

\subsection{Binaries Reported by the IAS Search}
We now focus on the 9 event candidates reported by the IAS search pipeline \cite{Venumadhav:2019tad,Zackay:2019tzo,Venumadhav:2019lyq,Zackay:2019btq}. GW151216 was first reported as a new, significant trigger in \cite{Zackay:2019btq} with $p_{\rm astro} \sim 0.71$ and a high, positive effective spin $\chi_{\rm eff} \sim 0.8$. This event was subsequently reported in \cite{2-OGC}, though with a markedly lower $p_{\rm astro} \sim 0.18$. In our Bayesian framework, we find that the probability that the signal is of astrophysical origin is overwhelmingly disfavoured, $\pastro \sim 4 \times 10^{-4}$, in broad agreement with the results presented in \cite{Ashton:2020odd}.

Of the six binaries reported in \cite{Venumadhav:2019lyq}, we find that \textit{three} of the event candidates, GW170121, GW170425 and GW170727, are of particular interest, with astrophysical probabilities $\pastro$ of $0.53$, $0.74$ and $0.66$ respectively. These three events are also the only candidates with $\log_{10} \mathcal{B}_{cr} > 0$. 

GW170121 is the most significant candidate reported in \cite{Venumadhav:2019lyq}, with both the highest SNR and a $p_{\rm astro} > 0.99$. This trigger was also subsequently found as a significant event candidate in \cite{2-OGC}, with $p_{\rm astro} \sim 1$. This event is notable due to its negative effective spin $\chi_{\rm eff} \approx -0.2$ \cite{Venumadhav:2019lyq}, see also Fig.~\ref{fig:corner_new_events_GW170121} in Appendix~\ref{app:posteriors}. 

GW170425 and GW170727 are consistent with being generated by heavy BBHs with source frame chirp masses $\mathcal{M} \approx 30 M_{\odot}$ and effective aligned spins $\chi_{\rm eff} \sim 0$, in broad agreement with the population of binaries observed in GWTC-1 \cite{LIGOScientific:2018jsj}. We report the measured parameters of these three events in Appendix~\ref{app:posteriors} and Figures~\ref{fig:corner_new_events_GW170121},~\ref{fig:corner_new_events_GW170425},~\ref{fig:corner_new_events_GW170727}, and \ref{fig:posterior_consistency}. 

Of the remaining candidates reported in \cite{Venumadhav:2019lyq}, we find that only GW170403 is of marginal interest, with an astrophysical probability $\pastro = 0.27$. The other two candidates, GW170202 and GW170304, have a $\pastro$ comparable to GW170818. 

The final two event candidates found in the IAS search, GWC170402 and GW170817A, were first reported in \cite{Zackay:2019btq}, a search targeting compact binary mergers that produce a clear signal in one of the detectors and a marginal signal in the other detectors. As per the discussion for GW170818, we do not expect the astrophysical signal probability returned by our framework to be large, finding $\pastro \sim 0$ and $0.02$ respectively. Unlike GW170818, the Bayesian coherence ratio also significantly disfavours these events, with $\log_{10} \mathcal{B}_{\rm cr}$ being more than an order of magnitude smaller. 

\subsection{Binaries Reported by the PyCBC Search}

As per the event candidates reported by the IAS search, we find that only the three events reported in the PyCBC catalogue~\cite{2-OGC}, GW170121, GW170425 and GW170727 are of interest. All other event candidates have negligible astrophysical probability. The PyCBC reported $p_{\rm astro}$ is $\sim 1$ for GW170121, $\sim 0.21$ for GW170425, and $\sim 0.99$ for GW170727. 

The other significant event reported in \cite{2-OGC} is GW170304, which was assigned $p_{\rm astro} = 0.7$ but for which we find that the astrophysical signal hypothesis is disfavoured with $\pastro = 0.03$. 

%%%%%%%%%%% FIGURE 
\begin{figure*}[ht!]
    \centering
    \includegraphics[width=\linewidth]{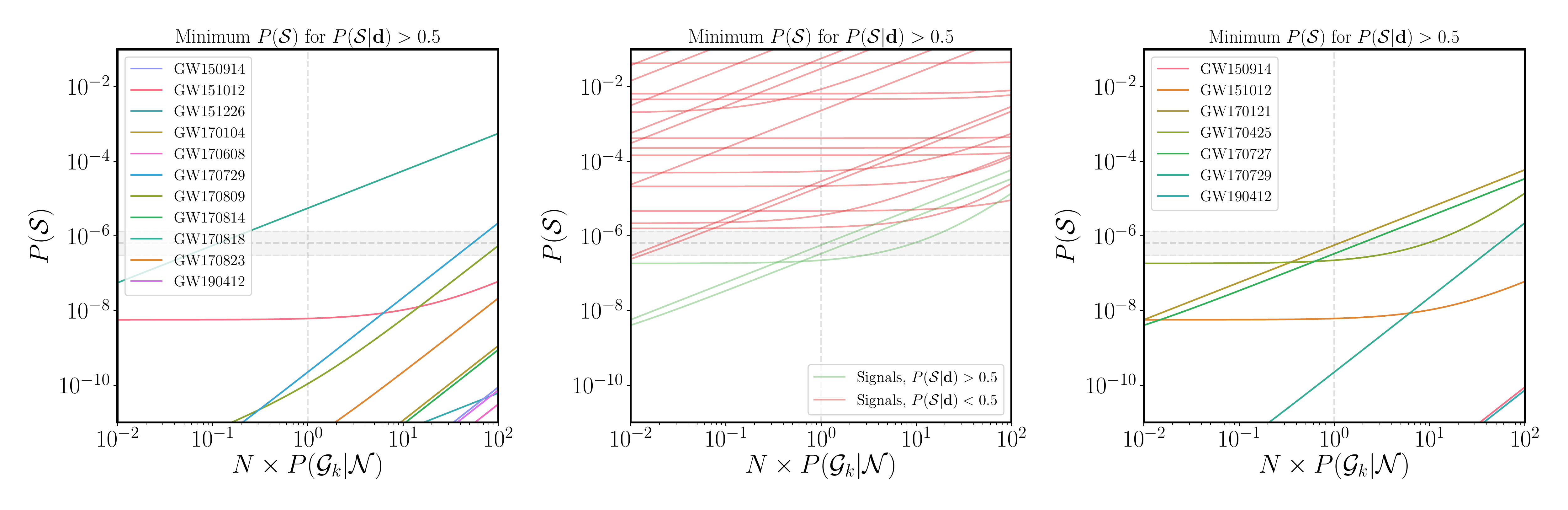}
        \caption{The minimum $P(\mathcal{S})$, which scales linearly with the total astrophysical merger rate integrated over the entire population, required to obtain a probability that the signal is of astrophysical origin $\pastro > 0.5$ as a function of the glitch probability $P(\mathcal{G}_k | \Hn)$. Here we scale the fiducial values $P(\Hg_k | \Hn)$ by a factor $N \in \left[ 10^{-2}, 10^{2} \right]$, keeping the ratio of the glitch probabilities between the detectors fixed, where the ratio is set by the values reported in Table~\ref{tab:PGk}. The left panel shows all confident events identified in GWTC-1 and GW190412. The middle panel shows all other triggers listed in Table.~\ref{tab:triggers}. The right panel shows a subset of interesting events, including the three most interesting triggers identified: GW170121, GW170425 and GW170727. The grey shaded regions denote the range of values $P(\Hs)$ can take by varying the inferred local merger rate $\mathcal{R}_0$ over its 90\% credible interval~ \cite{LIGOScientific:2018jsj}. The vertical dashed lines denote the fiducial values of $P(\mathcal{G}_k | \Hn)$.}
    \label{fig:alpha_beta}
\end{figure*}
%%%%%%%%%%

\subsection{Sensitivity of Results to Priors}
In Fig.~\ref{fig:PS_vs_PSd} we show how sensitive $P(\Hs | \dataall)$ is to the choice of the prior probability on the signal, $P(\Hs)$. Even if we reduce the value of $P(\Hs)$ by two orders of magnitude, we find that $P(\Hs | \dataall) \approx 1$ for all GWTC-1 BBHs, with the exception of GW151012, which would be recorded with  $P(\Hs | \dataall) \approx 0.6$, and GW170818, whose value would be $\approx 0$ as per the discussion above. 

The astrophysical signal probability of GW170121 and GW170425 is always above 0.5, even when taking the lowest value of $P(\Hs)$ compatible with the 90\% probability interval of the local BBH merger rate ${\cal R}_0$. For GW170727, the result drops just below our arbitrary threshold of $0.5$ to $P(\Hs | \dataall) \approx 0.4$. A signal prior $P(\Hs)$ that were at least an order of magnitude higher would yield $P(\Hs | \dataall) > 0.5$ for GW170403, GW170721 and GW170202. 

In Fig.~\ref{fig:PG_vs_PSd} we explore how sensitive the results are to the prior probability of the glitch hypothesis at a fixed signal probability corresponding to the value in Eq.~(\ref{eq:P_signal}). Again, all the GWTC-1 BBHs, with the usual exception of  GW170818, are fairly insensitive to variations in $P(\Hg_k | \Hn)$ by at least a factor of 100. 

For GW170425, one would need to increase the glitch probability by a factor $\approx 10$ to reduce $\pastro$ below 0.5. 

In particular GW170121, but also GW170727, are much more sensitive to variations in $P(\Hg_k | \Hn)$. Increasing the glitch prior $P(\Hg_k | \Hn)$ by a factor of $\approx 2-3$ would reduce $\pastro$ to below 0.5. 

Interestingly, $\pastro$ for GW170403, GW170721 and GW170202 is insensitive to the glitch probability in the region considered here: the signals are weak, and as discussed above $\pastro$ is driven by the prior signal probability. 

In Appendix~\ref{app:glitch_prob} and Fig.~\ref{fig:pastro_contours}, we provide additional diagnostics for a subset of events, including GW170425, GW170121 and GW170227, to show how $\pastro$ varies in the $P(\Hg_L | \Hn) - P(\Hg_H | \Hn)$ plane.

A complementary way of exploring how the astrophysical signal odds and associated signal posterior probability depend on the signal and glitch prior probabilities is displayed in Fig.~\ref{fig:alpha_beta}. We show the minimum $P(\Hs)$ required to obtain an astrophysical probability of $\pastro > 0.5$ as a function of the glitch probability $P(\Hg_k | \Hn)$, where the ratio of $P(\Hg_k | \Hn)$ between the detectors is kept constant, and set by the values reported in Table~\ref{tab:PGk}. The left panel of Fig.~\ref{fig:alpha_beta} shows all confident GWTC-1 events plus GW190412. The middle panel shows all remaining triggers, with the three most compelling events GW170121, GW170425 and GW170727 highlighted in green. The right panel of Fig.~\ref{fig:alpha_beta} focuses on a subset of interesting events. We show GW150914 and GW190412 as clear, unambiguous detections where even increasing the glitch rate by a factor $\sim \mathcal{O}(10^2)$ is insufficient to disfavour the astrophysical hypothesis. GW151012 was initially identified with a high false alarm rate (FAR) \cite{TheLIGOScientific:2016pea} but improvements to the analysis pipelines used for the GWTC-1 re-analysis of this event substantially reduced the FAR, leading to a $p_{\rm astro} = 0.96$ \cite{LIGOScientific:2018mvr}. In our Bayesian framework, GW151012 is unambiguously identified as an astrophysical signal with $\pastro \sim 0.99$, in agreement with \cite{Isi-et-al:2018,Ashton:2020odd} and showcasing the utility of our unified framework for calculating the astrophysical signal probability. Finally, we highlight the three most compelling events from our analysis presented: GW170121, GW170425 and GW170727. 

The horizontal shaded area in Fig.~\ref{fig:alpha_beta} denotes the allowed range of the prior probability on the signal as inferred from Eq.~(\ref{eq:p_S}) obtained by varying the local astrophysical merger rate $\mathcal{R}_0$ over its $90\%$ confidence interval. For all events with significant support for the astrophysical signal hypothesis, the minimum $P(\mathcal{S})$ required to meet the threshold of $0.5$ is far below the astrophysical rate estimated in App.~\ref{app:sig_rate} for typical glitch rates estimated from the distribution of \omicron triggers, with an upper limit on the glitch prior of $P(\Hg_k | \Hn) \sim \rm{few} \times 10^{-3}$. As an interesting case study, for GW170729 we would need to increase the glitch rate by a factor of $\sim 10^2$ before $\pastro$ starts to drop below $0.5$, emphasising the robustness of this result. 

\begin{figure*}
    \centering
    \includegraphics[width=0.9\linewidth]{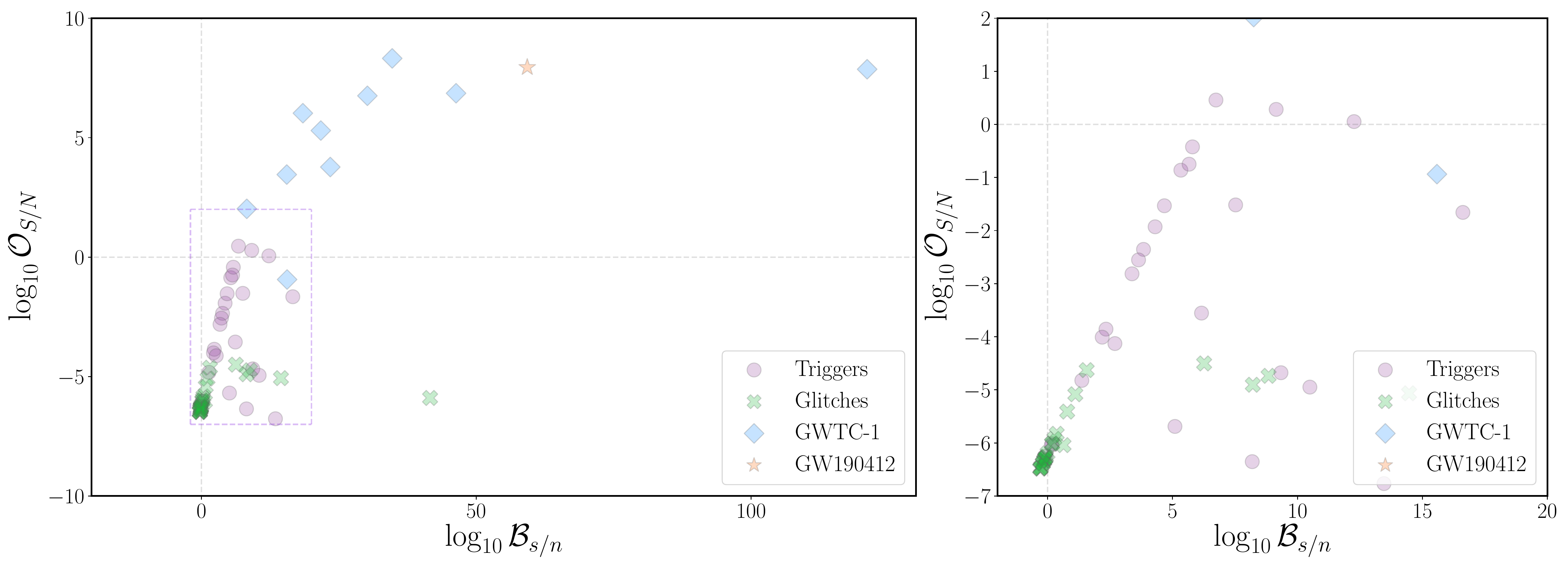}
        \caption{Bayes factor for the signal to noise hypothesis $\mathcal{B}_{s/n}$ against the astrophyiscal signal odds $\mathcal{O}_{S/N}$ for all events in GWTC-1 (blue), GW190412 (orange), the candidate events listed in Tab.~\ref{tab:triggers} (red) and a set of known glitches (green). Despite relatively high signal-to-noise Bayes factors, the astrophysical signal odds for the population of known glitches is efficiently suppressed within this framework. In the left panel we show all triggers and in the right panel we focus on the boxed region region demarcated by the purple dashed lines.}
    \label{fig:log10Bsn_vs_log10Osn}
\end{figure*}

It would be reasonable to wonder if our approach is robust against noise transients of instrumental origin that pollute the data. In Fig.~\ref{fig:log10Bsn_vs_log10Osn}, we show the Bayes factor for the signal to noise hypothesis $\mathcal{B}_{s/n}$ against the astrophysical signal odds $\mathcal{O}_{S/N}$ for all triggers listed in Tab.~\ref{tab:triggers}. In addition, we plot the same quantities evaluated for a population of known glitches identified by the PyCBC search. Figure~\ref{fig:log10Bsn_vs_log10Osn} serves to demonstrate the efficacy with which the inferred $\pastro$ for glitches is suppressed and as a caution that the standard Bayes factor for signal vs noise $\mathcal{B}_{s/n}$ is a poor discriminator between astrophysical signals and instrumental noise transients.

%%%%%
\begin{figure*}[ht!]
    \centering
    \includegraphics[width=\linewidth]{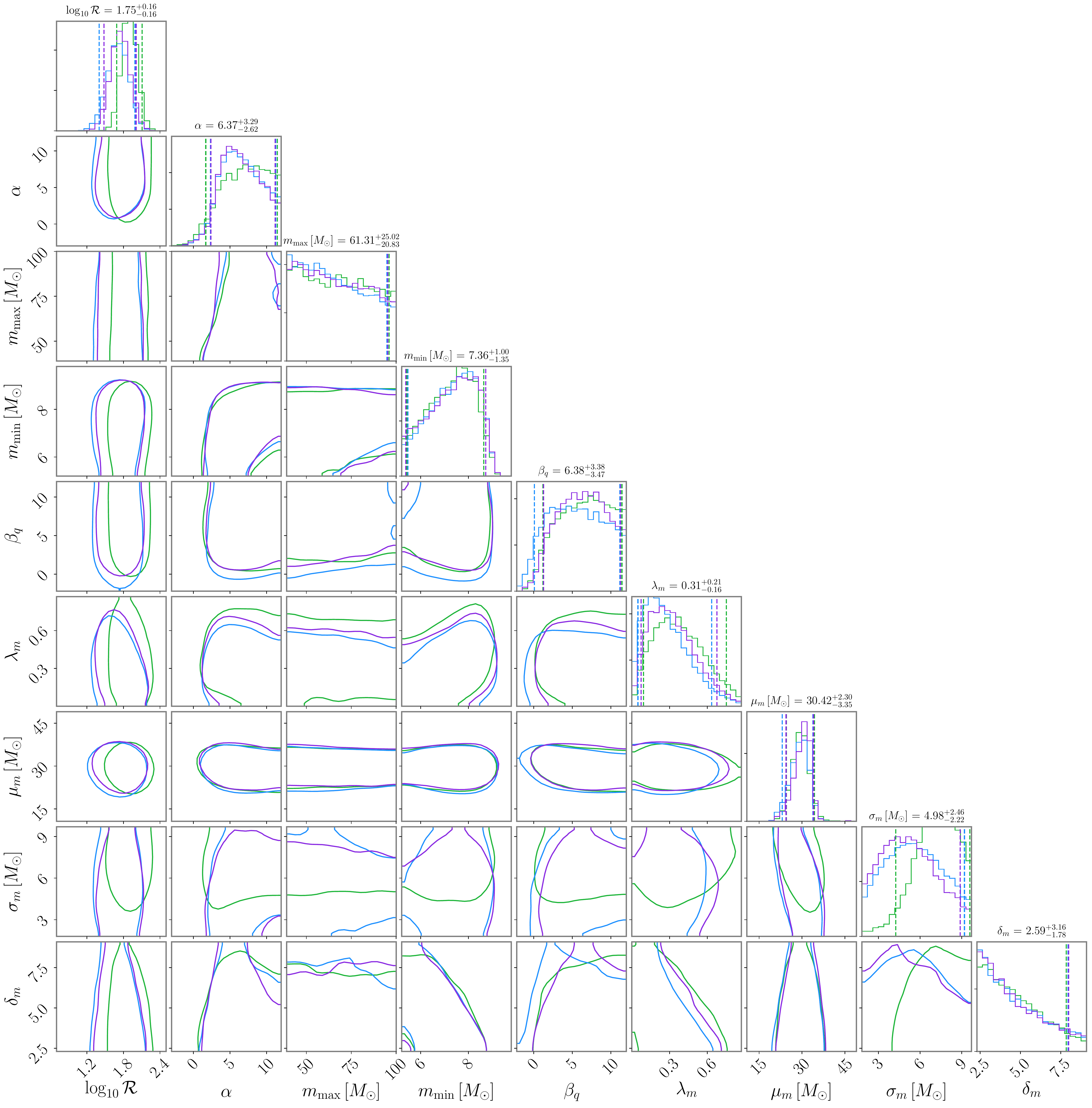}
        \caption{The 1D and 2D posterior distributions for the hyperparameters describing the BBH population model for the masses and rates. Here the blue posteriors denote the results of the analysis considering all GWTC-1 events in the population analysis. The red posteriors show the results when including the three events with $\pastro > 0.5$, GW170121, GW170425 and GW170727 in addition to all GWTC-1 events. The green posteriors refer to the inferred population based on all GWTC-1 events plus all triggers reported in \cite{Zackay:2019btq,Zackay:2019kkv,Zackay:2019tzo,Venumadhav:2019tad} irrespective of $\pastro$. We also report the 1D median values and 90\% probability interval inferred using all GWTC-1 binaries plus the three events with $\pastro > 0.5$. }
    \label{fig:poppe}
\end{figure*}
%%%%%

\section{Implications for Population Inference}
\label{sec:poppe}

In our Bayesian analysis, we have identified three additional event candidates with a $\pastro > 0.5$ (see Table~\ref{tab:triggers}), which would be sufficient for inclusion in typical population studies \cite{LIGOScientific:2018mvr,LIGOScientific:2018jsj}. In this section, we assess the impact of including GW170121, GW170425 and GW170727 on the inferred BBH population properties~\cite{LIGOScientific:2018jsj}. 

Following the methodology described in~\cite{LIGOScientific:2018jsj}, we perform hierarchical Bayesian inference, incorporating measurement uncertainty and selection effects into the analysis \cite{TheLIGOScientific:2016htt,Wysocki:2018mpo,Mandel:2018mve,Fishbach:2018edt}. The rate of events is modeled as a Poisson process whose mean is dependent on the parameter distribution of the population of binary black holes 
\begin{align}
\mathcal{L}( \lbrace d_n \rbrace | \Lambda ) &\propto e^{- \mu (\Lambda)} \displaystyle \prod^{N_{\rm obs}}_{n = 1} \mathcal{L} (d_n | \theta , z) \frac{d N}{d \theta d z} (\Lambda) \, d \theta \, d z ,
\end{align}
where $\theta$ are the intrinsic parameters of the binary, $z$ the redshift, $N$ the total number of mergers that occur within the detection horizon, and $\mu (\Lambda)$ the rate constant describing the mean number of events as a function of the population hyperparameters $\Lambda$. Here $N_{\rm obs}$ is the number of detections and the individual-event likelihood for the $n^{\rm th}$ detection is $\mathcal{L}(d_n | \theta , z)$. The expected number of events is dependent on the selection effects for the detectors. This can be characterised by the sensitive spacetime volume $VT (\theta)$ for a detector network in terms of the detection probability for a binary with parameters $\theta$ and the total observation time, as discussed in App.~\ref{app:sig_rate}.  

%%%%%
\begin{figure*}
    \centering
    \includegraphics[width=0.65\linewidth]{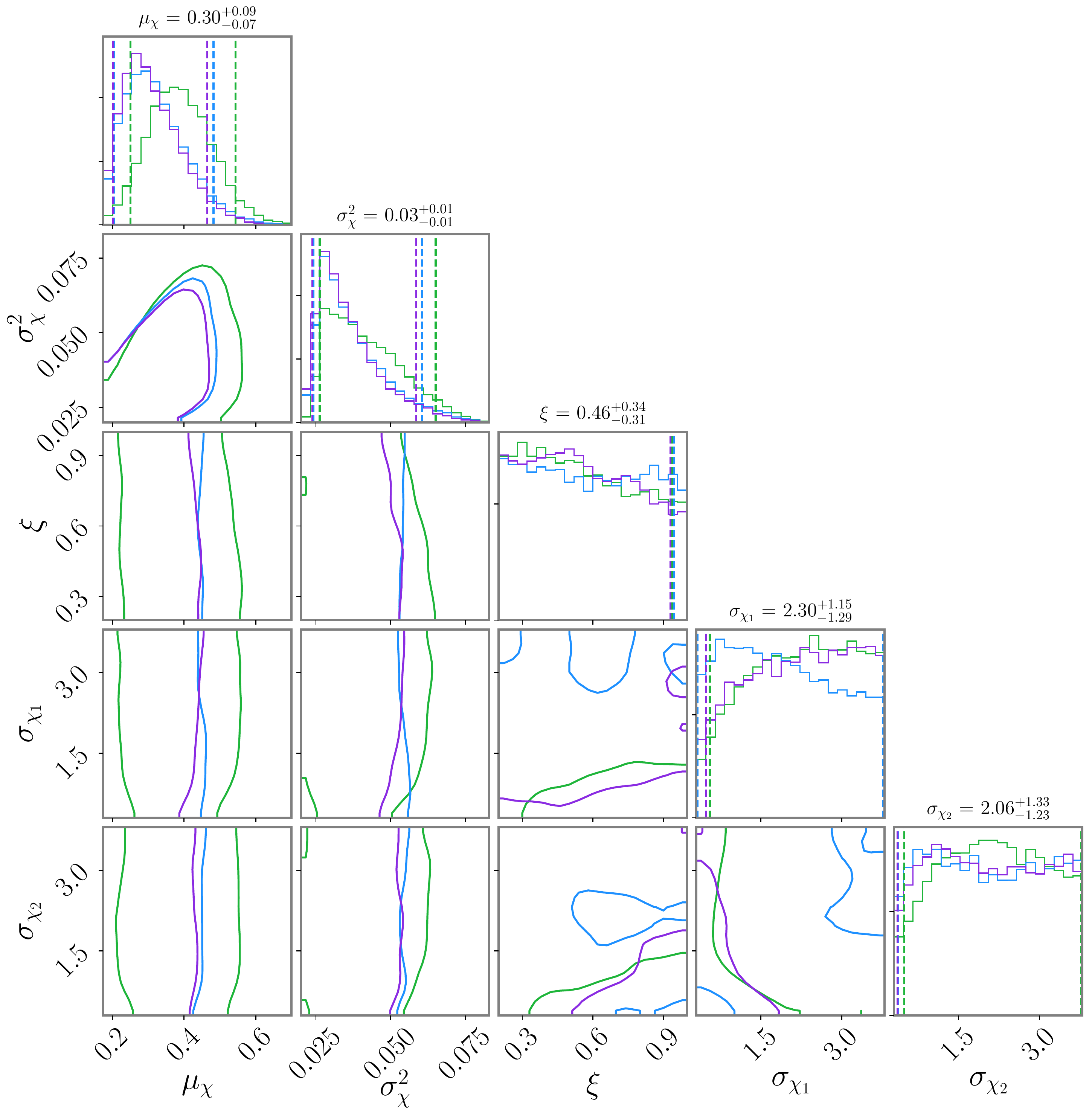}
        \caption{The 1D and 2D posterior distributions for the hyperparameters describing the spin population model. As per Fig.~\ref{fig:poppe}, the blue posteriors denotes all GWTC-1 events. The red posteriors show all GWTC-1 events plus the three events with $\pastro > 0.5$, GW170121, GW170425 and GW170727. The green posteriors show all GWTC-1 events plus all triggers reported in \cite{Zackay:2019btq,Zackay:2019kkv,Zackay:2019tzo,Venumadhav:2019tad}. We report the 1D median values and 90\% probability interval inferred using all GWTC-1 binaries plus the three events with $\pastro > 0.5$.}
    \label{fig:poppe_s}
\end{figure*}
%%%%%

For the population inference, we use \textsf{Model C} of \cite{LIGOScientific:2018jsj} as a representative model. This model is based on a power law mass distribution with an additional Gaussian component at high masses \cite{Talbot:2018cva}. The low-mass cutoff $m_{\rm min}$ is tapered by a smoothing scale $\delta m$ to account for environmental effects, such as metallicity, that can blur the edge of the lower mass gap \cite{Talbot:2018cva,LIGOScientific:2018jsj} and the maximum allowed BH mass is given by $m_{\rm max}$. At low masses, we recover a standard mass power-law governed by a power-law index $\alpha$ on the primary BH and a mass ratio power-law index $\beta_q$. The Gaussian component models the possible build up of high-mass BBHs from pulsational pair-instability supernovae \cite{Barkat:1967zz,Woosley:1986pisn,Heger:2001cd,Chatzopoulos:2012pisn,Woosley:2016hmi} and is parameterized by the mean $\mu_m$ and standard deviation $\sigma_m$ and well as the fraction of primary BHs in the Gaussian component $\lambda_m$. The distribution of spin magnitudes $\chi_i$ is taken to be drawn from a beta distribution \cite{Wysocki:2018mpo} and can be parameterized by either the moments of the distribution $\alpha_{\chi}$ and $\beta_{\chi}$ or, equivalently, by the mean $\mu_{\chi}$ and variance $\sigma^2_{\chi}$. The distribution of spin orientations is modelled as a mixture of two distributions corresponding to an isotropic and a preferentially aligned component \cite{Talbot:2017yur}. The fraction of binaries preferentially aligned with the orbital angular momentum is denoted $\xi$ and the degree of spin misalignment is denoted by $\sigma_{\chi_i}$. The redshift evolution of the model follows the prescription in \cite{Fishbach:2018edt}, where, for simplicity, we assume a merger rate density that is uniform in comoving volume and source-frame time. 

The likelihood $\mathcal{L}(d_n | \theta, z)$ is recycled from the posterior samples calculated using \texttt{LALInference} \cite{Veitch:2014wba}. We assume uniform priors on the population parameters, as detailed in Table 2 of \cite{LIGOScientific:2018jsj}. For the event rate, we take a log-uniform distribution bounded between $\left[ 10^{-1} , 10^3 \right]$. The hierarchical population inference is carried out using the \texttt{gwpopulation} package \cite{Talbot:2019okv} and \texttt{Bilby} \cite{Ashton:2018jfp}.  

As a proxy for incorporating full information regarding the astrophysical signal probability, we use the posterior probability of the signal hypothesis $P(\Hs | \dataall)$ as a criterion for including events in our population inference. Following \cite{LIGOScientific:2018mvr,LIGOScientific:2018jsj}, the threshold is taken to be $P(\Hs | \dataall) > 0.5$. Based on the analysis discussed in Sec.~\ref{sec:results}, the events GW170121, GW170425 and GW170727 satisfy this selection criterion for our default priors and we can incorporate them into the hierarchical inference. 

Some approaches for incorporating $p_{\rm astro}$ in hierarchical population inference have recently been presented in \cite{Gaebel:2018poe,Galaudage:2019jdx}. The framework presented in \cite{Galaudage:2019jdx} also allows for $p_{\rm astro}$ to be revised by taking into account information regarding the distribution of black hole masses and spins from the observed population of binaries. Applying this method to all GWTC-1 candidates, as well as the 8 binaries reported in \cite{Zackay:2019tzo,Venumadhav:2019lyq,Venumadhav:2019tad}, \cite{Galaudage:2019jdx} determine that all binaries have a $p_{\rm astro} \sim 1$. This places the results in some tension with the results presented here and in \cite{Ashton:2020odd}. Whilst a detailed re-analysis of the impact of $p_{\rm astro}$ on population inference is beyond the scope of this paper, it seems plausible that discarding information regarding the coherence of the signal is in part culpable for the discrepancies observed and could lead to a non-trivial bias in the inferred astrophysical merger rate. 

In Figures~\ref{fig:poppe} and~\ref{fig:poppe_s} we compare the posterior distributions for the mass and spin population hyperparameters, respectively, by including in the population study: (i) GWTC-1 binaries alone (blue curves), (ii) GWTC-1 binaries and the three new event candidates with significant astrophysical probability as determined by our analysis (purple curves), and (iii) GWTC-1 binaries and \emph{all} binaries reported in the IAS search, excluding GW170402, irrespective of $\pastro$ as in~\cite{Galaudage:2019jdx}.
The population hyperparameters inferred from GWTC-1 plus the three new events are in broad agreement with the results inferred using the GWTC-1 binaries alone. We see a slight shift to higher values of the astrophysical merger rate $\mathcal{R}_0$ and some of the other hyperparameters, notably the degree of spin misalignment for the primary BH $\sigma_{\chi_1}$. 

Incorporating the 8 IAS event candidates, we find that the population hyper-parameters are increasingly discrepant with respect the results from using GWTC-1 alone. In particular, we observe that the astrophysical merger rate $\mathcal{R}_0$, the width of the Gaussian mass peak $\sigma_m$ and the mean of the spin magnitude distribution $\mu_{\chi}$ shift quite significantly. We observe the same shift in $\sigma_{\chi_1}$ as when including the 3 events with significant astrophysical probability returned by our analysis. 

Incorporating information regarding the coherence of the signal will likely be an important consideration when performing population inference on an ever growing catalog of observed compact binaries. This highlights the necessity of a robust, unified framework for calculating the astrophysical signal odds agnostic to any specific search pipeline of strategy, such as the framework presented here.

%%%%%%%%%%%%%%% DISCUSSION & CONCLUSIONS
\section{Discussion}
\label{sec:discussion}
We have revisited the problem of determining whether an observed GW transient is due to an astrophysical source. We employ a Bayesian framework to derive the posterior odds ratio, $\mathcal{O}_{S/N}$, allowing us to determine the probability that the signal is of astrophysical origin, $\pastro$, independent of any search pipeline. The astrophysical posterior odds is determined through a coherent analysis of the data, depending only on the underlying model hypotheses and the choice of prior probabilities. In order to set the astrophysical signal prior $P(\Hs)$, we leveraged prior knowledge on the population of astrophysical BBHs. The glitch probability for a given detector $P(\Hg_k | \dataall)$ was determined using the density of \omicron triggers in a 24h stretch of data around each event, allowing us to account for asymmetric noise profiles between the detectors and for changes in the behaviour of the detectors over time. This work builds on recent studies in the literature \cite{VeitchVecchio:2010, Isi-et-al:2018, Ashton:2018jfp, AshtonThraneSmith:2019}.  

We analyse all confident BBH events reported in GWTC-1 \cite{LIGOScientific:2018mvr} as well as event candidates recently reported by independent search pipelines \cite{Venumadhav:2019lyq,Zackay:2019tzo,Zackay:2019btq,2-OGC,Nitz:2020naa}. Using the astrophysical posterior odds $\mathcal{O}_{S/N}$, we find that all GWTC-1 binaries, except for the single detector event GW170818, have overwhelming odds in favour of the astrophysical signal hypothesis. Of all other event candidates analyzed, only GW170121, GW170425 and GW170727 have odds in favour of the astrophysical signal hypothesis, $\pastro > 0.5$, meeting the threshold for inclusion in typical population studies \cite{LIGOScientific:2018jsj}. We characterized the robustness of these results to changes in the signal $P(\Hs)$ and glitch $P(\Hg_k | \Hn)$ priors. 
Due to the significant computational burden of our analysis, in this paper we have considered only BBH candidates; our method however is fully general and can be applied to all classes of binary systems.

The methodology presented in this paper can be further improved in a number of ways. Due to the limited coincident data available, and the non-trivial computational processing load we restricted our analysis to a 2-detector network. As the duty cycle of the instruments improves, the analysis should incorporate data from all available detectors in a given network at any given time. The fourth observing run (O4) is scheduled to last for one year, with the LIGO detectors nearing design sensitivity and Phase 1 of the Advanced VIRGO+ upgrade nearing completion \cite{Aasi:2013wya}. In addition, KAGRA \cite{Aso:2013eba,Akutsu:2018axf} will be operational with a nominal BNS range of $25-130 \rm{Mpc}$ \cite{Aasi:2013wya}, providing the potential opportunity for analysing coherent data from 4 detectors. The discriminating power of our approach between astrophysical signals and instrumental transient will correspondingly increase as data from more instruments are included. 

With detector sensitivities ever increasing, the approximation that signals and glitches do not overlap can break down, as was the case for GW170817 \cite{TheLIGOScientific:2017qsa}. Whilst glitch subtraction can be an effective tool to mitigate such scenarios \cite{Cornish:2014kda}, the model will need to be generalised to include this possibility, and in the future even contemplate the possibility of overlapping astrophysical signals.

For computational efficiency, we used the precessing waveform model IMRPhenomPv2 \cite{Hannam:2013oca,Schmidt:2014iyl,Husa:2015iqa,Khan:2015jqa}. In future analyses, we plan to utilise improved and more accurate waveform models \cite{Pratten:2020fqn} that incorporate both higher modes \cite{Nagar:2019wds,Nagar:2020pcj,Garcia-Quiros:2020qpx} and precession \cite{Pratten:2020ceb,Ossokine:2020kjp}. As a first step, we have re-analysed GW170121, GW170425 and GW170727 with the precessing higher-mode waveform model PhenomXPHM \cite{Pratten:2020ceb}, while keeping all the other parameters of our analysis unchanged.
We find that $\pastro$ increases from $0.53$ to $0.56$ for GW170121, from $0.74$ to $0.84$ for GW170425, and from $0.66$ to $0.67$ for GW170727.

To process large catalogues of candidate events or inferring $\pastro$ in low-latency -- the latter can be done with rather minor changes to the present implemented software infrastructure and relatively small additional computational cost -- it will be important to adopt methods that further reduce waveform generation costs \cite{Vinciguerra:2017ngf,Garcia-Quiros:2020qlt} and mitigate the computational cost of the Bayesian inference \cite{Canizares:2014fya,Smith:2016qas}. 

Due to the lack of a robust glitch model, we made the simplifying approximation that $\Hg_k \approx \Hs_k$. Whilst this is the most conservative choice that could be made \cite{VeitchVecchio:2010}, our understanding of the instruments' behaviour is continuously improving, as is our ability to classify and describe glitches. In the future, a reliable glitch model could be available and allow us to more adequately distinguish between astrophysical signals and glitches, thereby increasing the significance of the inferred $\pastro$. 

The astrophysical signal prior $P(\Hs)$ used in this analysis was determined by calculating the expected number of observed BBHs within the relevant time span from a population of binaries taking into account selection effects, see Eq.~(\ref{eq:num_det}). A more detailed characterization of the detector sensitivity $\langle VT \rangle_{\Lambda}$, e.g. through Monte-Carlo integration \cite{Tiwari:2017ndi,LIGOScientific:2018jsj} or novel applications of machine-learning \cite{Gerosa:2020pgy}, could allow for a more accurate determination of $P(\Hs)$ as a function of the population hyperparameters $\Lambda$. In addition our analysis can be improved by dividing the whole observing period into shorter segments, and updating $P(\Hs)$ as the analysis progresses. Finally, the default priors used in calculating the Bayes factors are uniform in the component masses, uniform in spin magnitudes and isotropic in spin orientations. The use of such agnostic priors is of particular importance in the absence of any a priori knowledge about the expected shape of the distributions. However, as the number of observations increases, it will be important to consider population informed priors, e.g. by constructing the posterior predictive distributions \cite{Fishbach:2019ckx,Galaudage:2019jdx}. 

%%%%%%%%%%%%%%%%%%%%%%%%%%%%
\section*{Acknowledgments}
We thank Patricia Schmidt, Thomas Dent and Christopher Moore for useful discussions and comments on the manuscript. GP and AV acknowledge support from Science and Technology Facilities Council (STFC) Grant No.  ST/N021702/1. AV acknowledges the support of the Royal Society and Wolfson Foundation. This research has made use of data, software and/or web tools obtained from the Gravitational Wave Open Science Center (https://www.gw-openscience.org), a service of LIGO Laboratory, the LIGO Scientific Collaboration and the Virgo Collaboration. LIGO is funded by the U.S. National Science Foundation. Virgo is funded by the French Centre National de Recherche Scientifique (CNRS), the Italian Istituto Nazionale della Fisica Nucleare (INFN) and the Dutch Nikhef, with contributions by Polish and Hungarian institutes. We are grateful for computational resources funded by STFC grant ST/I006285/1 and ST/V001167/1 for Advanced LIGO Operations Support. This analysis made use of the CIT cluster provided by LIGO Laboratory and supported by National Science Foundation Grants PHY-0757058 and PHY-0823459.

%%%%%%%%%%%%%%% APPENDICES
\appendix

\section{Calculation of the signal prior probability}
\label{app:sig_rate}
Here we provide details of the calculation of the prior signal probability, $P(\Hs)$, given by Eq.~(\ref{eq:p_S}). 

In order to calculate the expected number of detections $\mu (\Lambda)$, we need to incorporate selection effects from a network of detectors. Here, we calculate the sensitive spacetime volume $\langle VT \rangle$ following a semi-analytic prescription \cite{Finn:1992xs,TheLIGOScientific:2016pea,LIGOScientific:2018jsj}. For a network of gravitational wave detectors, the sensitive spacetime volume is 
\begin{align}
    VT (\theta) &= T_{\rm obs} \displaystyle\int^{z_{\rm max}}_0 f(z | \theta) \frac{d V_c}{dz} \frac{1}{1+z} \, dz 
\end{align}
where we assume a constant sensitivity over an observing run $T_{\rm obs}$ and $f(z|\theta)$ is the detection probability for a binary with parameters $\theta$ at a redshift $z$ \cite{OShaughnessy:2009szr} averaged over extrinsic parameters \cite{Finn:1992xs}. For each binary, we calculate the optimal SNR $\rho_{\rm opt}$ using the IMRPhenomXAS phenomenological waveform model \cite{Pratten:2020fqn}, corresponding to the SNR that would be observed for a face-on source located directly overhead a detector. For an isotropic distribution of sources, with arbitrary sky locations and inclinations, the distribution of SNRs can be captured by introducing a coefficient $\Theta$ that parameterizes the angular dependence, $0 \leq \Theta = \rho / \rho_{\rm opt}$, and where $\Theta$ has a known distribution \cite{Finn:1992xs}. Here $\Theta = 1$ corresponds to an optimally oriented source and $\Theta = 0$ corresponds to a binary that is in a blind spot of the detector. The probability of detecting a source at a redshift $z$ can therefore be written as
\begin{align}
    f (z | \theta) &\equiv p_{\rm det} (z | \theta) = P (\rho \geq \rho_{\rm th}) ,
\end{align}
where $\rho_{\rm th}$ is the threshold for observing a binary. Here we assume that sources will only be detected if they have a single detector SNR above a threshold $\rho_{\rm ifo} > 7$. 

We incorporate both mass and spin dependence in the calculation of $VT (\theta)$, though spins have a subdominant effect on the $\rho$ calculation for the population of binaries considered here \cite{LIGOScientific:2018jsj}. For a given population with hyper-parameters $\Lambda$, the total sensitive spacetime volume is \cite{LIGOScientific:2018jsj}
\begin{align}
    \langle VT \rangle_{\Lambda} &= \displaystyle\int_{\theta} p (\theta | \Lambda) \, VT (\theta) \, d \theta .
\end{align}
If the merger rate evolves with redshift, the expected number of detections will be given by
\begin{align}
\label{eq:num_det}
\mu(\Lambda) &= T_{\mathrm{obs}} \int_{\theta} \int_{0}^{\infty} p(\theta | \Lambda) f(z | \theta) \mathcal{R} (z) \frac{\mathrm{d} V_{c}}{\mathrm{d} z} \frac{1}{1+z} \mathrm{d} z \, \mathrm{d} \theta .
\end{align}
Following \cite{Fishbach:2018edt}, we can parameterize the evolving merger rate density $\mathcal{R}(z)$ in the comoving frame by
\begin{align}
    \mathcal{R} (z | \lambda) &= \mathcal{R}_0 (1 + z)^{\lambda} ,
\end{align}
where $\mathcal{R}_0$ is the local merger rate density at $z = 0$. Here, $\lambda = 0$ corresponds to a uniform in comoving volume merger rate and $\lambda \sim 3$ is a merger rate that approximately follows the star formation rate \cite{Madau:2014bja}, at redshifts below $z\approx 2$, which corresponds to the peak of the star formation rate. For the population of binaries here, we take $\lambda = 2$, consistent with~\cite{LIGOScientific:2018jsj,LIGOScientific:2020stg}. In the simpler scenario in which the merger rate is constant with redshift, the expected number of observations reduces to
\begin{align}
    \mu(\Lambda) &= \mathcal{R}_0 \, \langle VT  \rangle_{\Lambda} .
\end{align}
The expectation value of $VT(\theta)$ can be calculated using standard Monte Carlo methods
\begin{align}
    \langle VT \rangle_{\Lambda} &= \displaystyle\int p(\theta | \Lambda) \, VT(\theta) \, d \theta 
    \approx \frac{1}{N} \displaystyle\sum^N_k VT (\theta_k) .
\end{align}
We use PSDs that are representative of the detector performance in O2 \cite{LIGOScientific:2018mvr,GWOSC} and assume a total coincident observing time of $T_{\rm obs} = 0.46 \rm{y}$ \cite{LIGOScientific:2018mvr}. For the population of binaries, we use a parameterized mass model 
\begin{align}
    p(m_1,m_2 | m_{\rm min}, m_{\rm max}, \alpha, \beta_q) \propto C(m_1) m_1^{-\alpha} \, q^{\beta_q} ,
\end{align}
where the minimum and maximum black hole masses are taken to be $m_{\rm min} = 5 M_{\odot}$ and $m_{\rm max} = 50 M_{\odot}$, as motivated by the apparent lower limit for X-ray binary observations \cite{Farr:2011} and the maximum mass above which a pair-instability supernova is thought to completely disrupt the star \cite{Barkat:1967zz}. The slope of the mass power law is taken to be $\alpha = 1.8$ and the exponent for the mass ratio to be $\beta_q = 2.0$, in broad agreement with \cite{LIGOScientific:2018jsj,LIGOScientific:2020stg}. The distribution of spin magnitudes $\chi_i$ are taken to follow a beta distribution \cite{Wysocki:2018mpo}
\begin{align}
    p\left(\chi_{i} | \alpha_{\chi}, \beta_{\chi}\right)=\frac{\chi_{i}^{\alpha_{\chi}-1}\left(1 - \chi_{i}\right)^{\beta_{\chi}-1}}{\mathrm{B}\left(\alpha_{\chi}, \beta_{\chi}\right)} ,
\end{align}
with $\alpha_{\chi} = 1.5$ and $\beta_{\chi} = 3.5$. The spin orientations are taken to follow a mixture model consisting of an isotropic component and a preferentially aligned component \cite{Talbot:2017yur}
\begin{align}
p\left(\cos t_{1}, \cos t_{2} | \sigma_{1}, \sigma_{2}, \zeta\right) &= \frac{(1-\zeta)}{4} \nonumber \\
+ \frac{2 \zeta}{\pi} \prod_{i \in\{1,2\}} & \frac{\exp \left(-\left(1-\cos t_{i}\right)^{2} /\left(2 \sigma_{i}^{2}\right)\right)}{\sigma_{i} \operatorname{erf}\left(\sqrt{2} / \sigma_{i}\right)},
\end{align} 
where we take $\zeta = 0.5$ and $\sigma_i = 1$. In practice, we only use the spins projected along the orbital angular momentum $\vec{L}$ when evaluating the aligned-spin waveform model to calculate the SNR.

Using the population of binaries detailed above, the sensitive spacetime volume is estimated to be
\begin{align}
    \langle VT \rangle_{\Lambda} &\approx 0.877 \; \rm{Gpc}^3 \; \rm{y} .
\end{align}
Using a fiducial local astrophysical merger rate of $\mathcal{R}_0 = 53.2 \, \rm{Gpc}^{-3} \, \rm{y}^{-1}$ \cite{LIGOScientific:2018jsj}, the probability of a signal within a coalescence interval of $0.2s$ yields Eq.~(\ref{eq:p_S}).
%
%can therefore be expressed as 
%\begin{align}
%    \alpha \approx 6.43 \times 10^{-7} %\left( \frac{ \mathcal{R}_0 }{53.2 \, %\rm{Gpc}^{-3} \, \rm{y}^{-1}} \right) \left( %\frac{\Delta t_c}{0.2 s} \right) .
%\end{align}

%%%%%%%%%%%%%%%%%%%%%%%%%%%%%%

\section{Impact of glitch prior probability on the posterior signal probability of a signal}
\label{app:glitch_prob}

In Table~\ref{tab:PGk} we report the glitch probability for each of the two instruments that we have used in the analysis, see Tab.~\ref{tab:triggers}, and we also report the SNRs (signle detector and across the network) for each of the candidate events.

\begin{figure*}[!htpb]
    \centering
    \includegraphics[width=0.8\columnwidth]{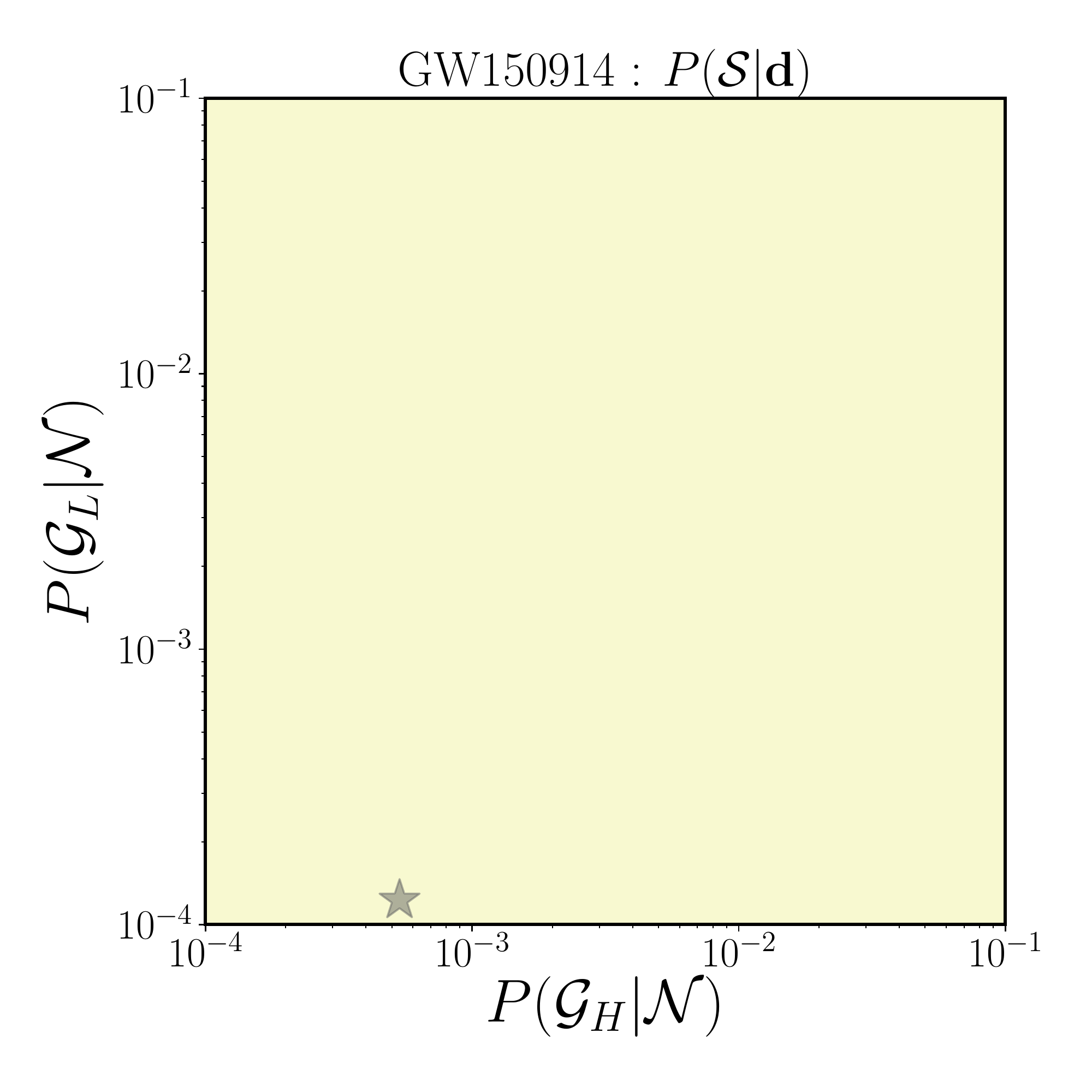}
    \includegraphics[width=0.8\columnwidth]{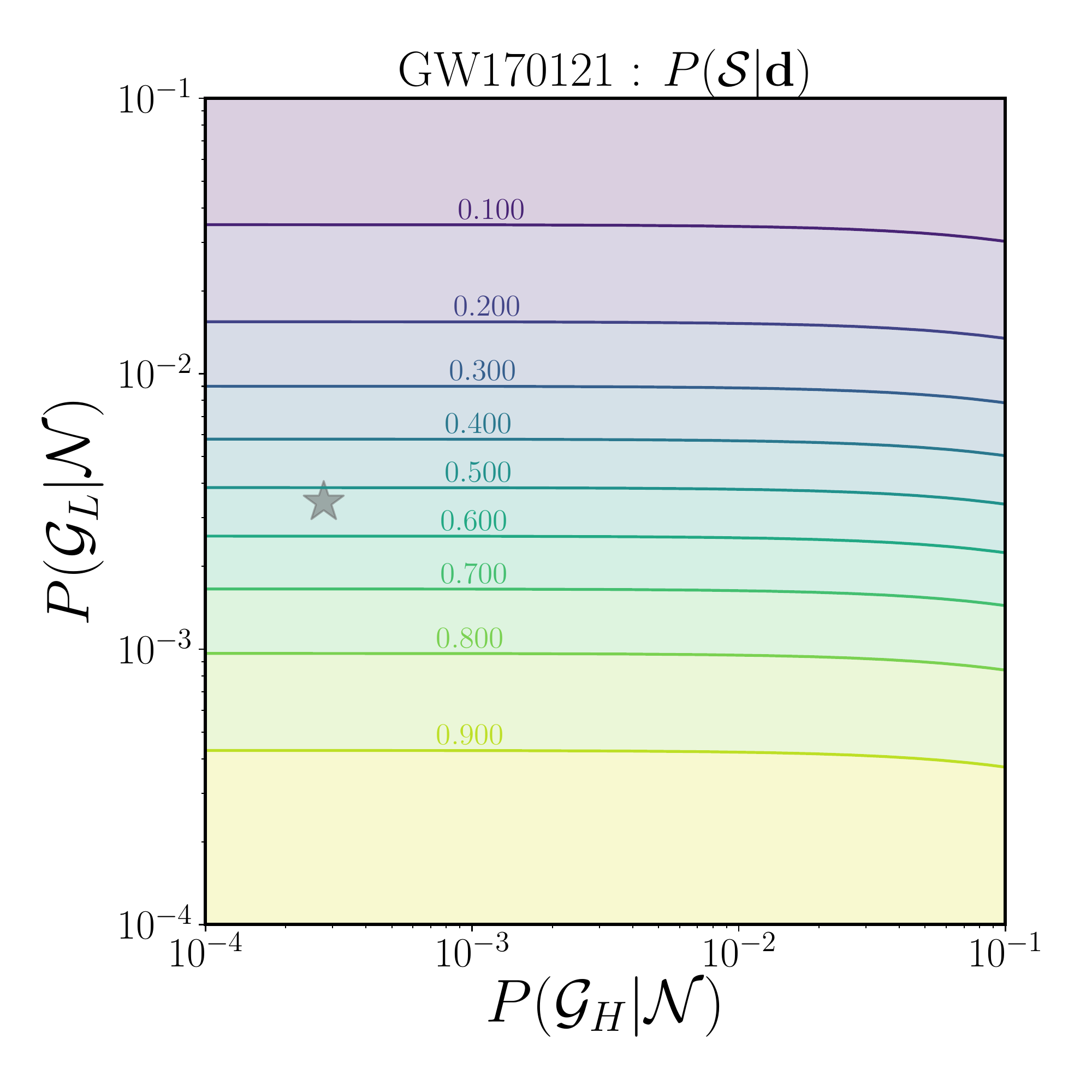}
    \includegraphics[width=0.8\columnwidth]{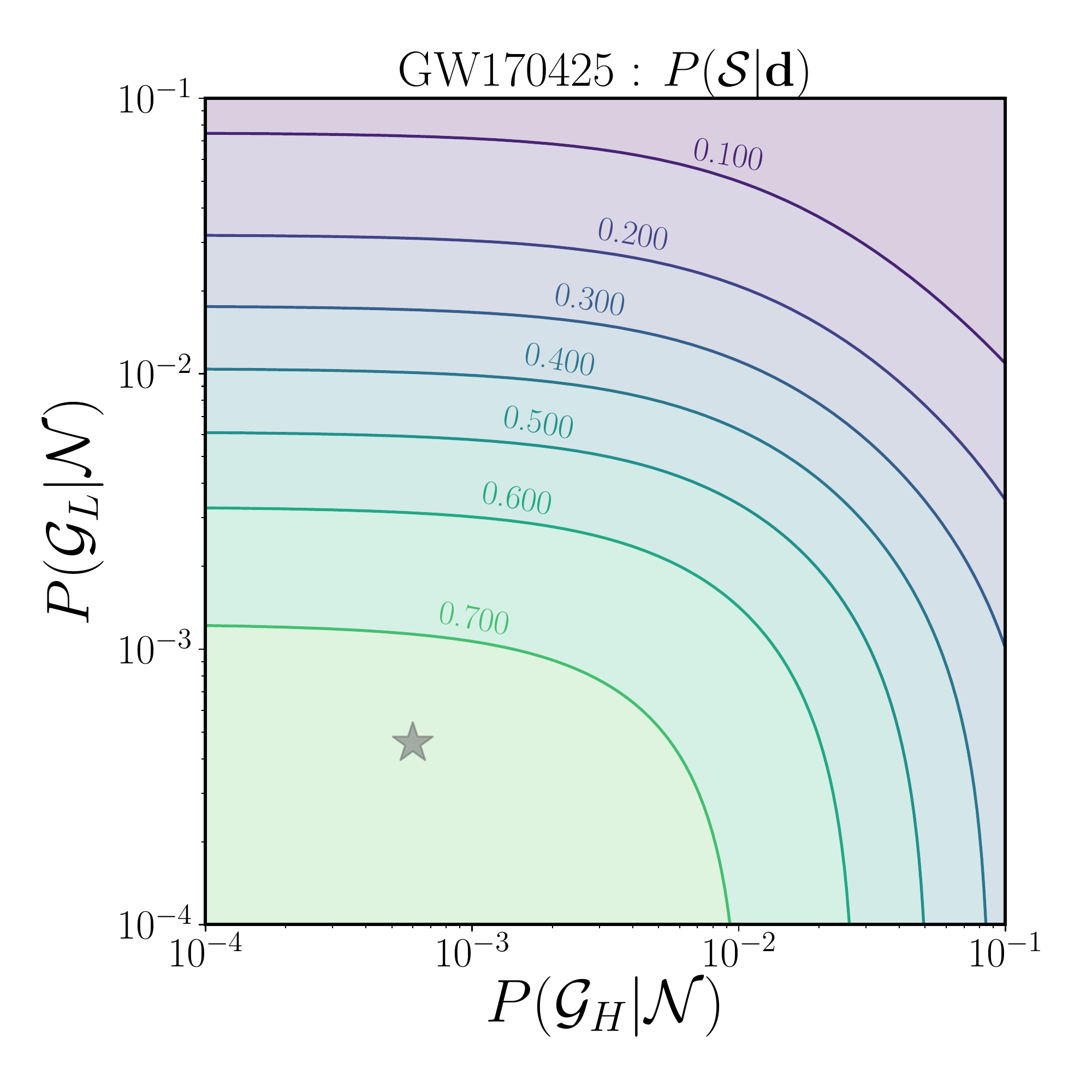}
    \includegraphics[width=0.8\columnwidth]{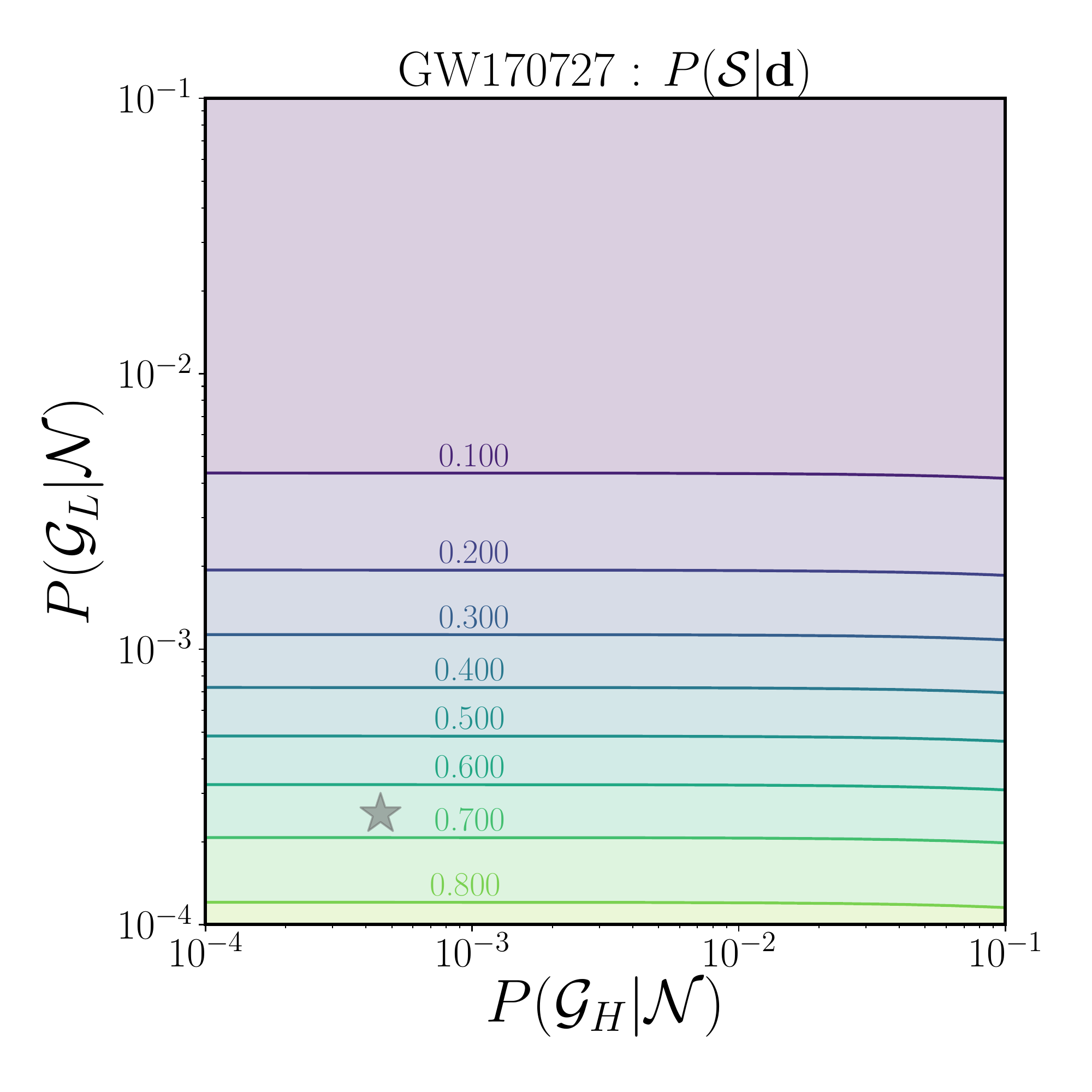}
    \includegraphics[width=0.8\columnwidth]{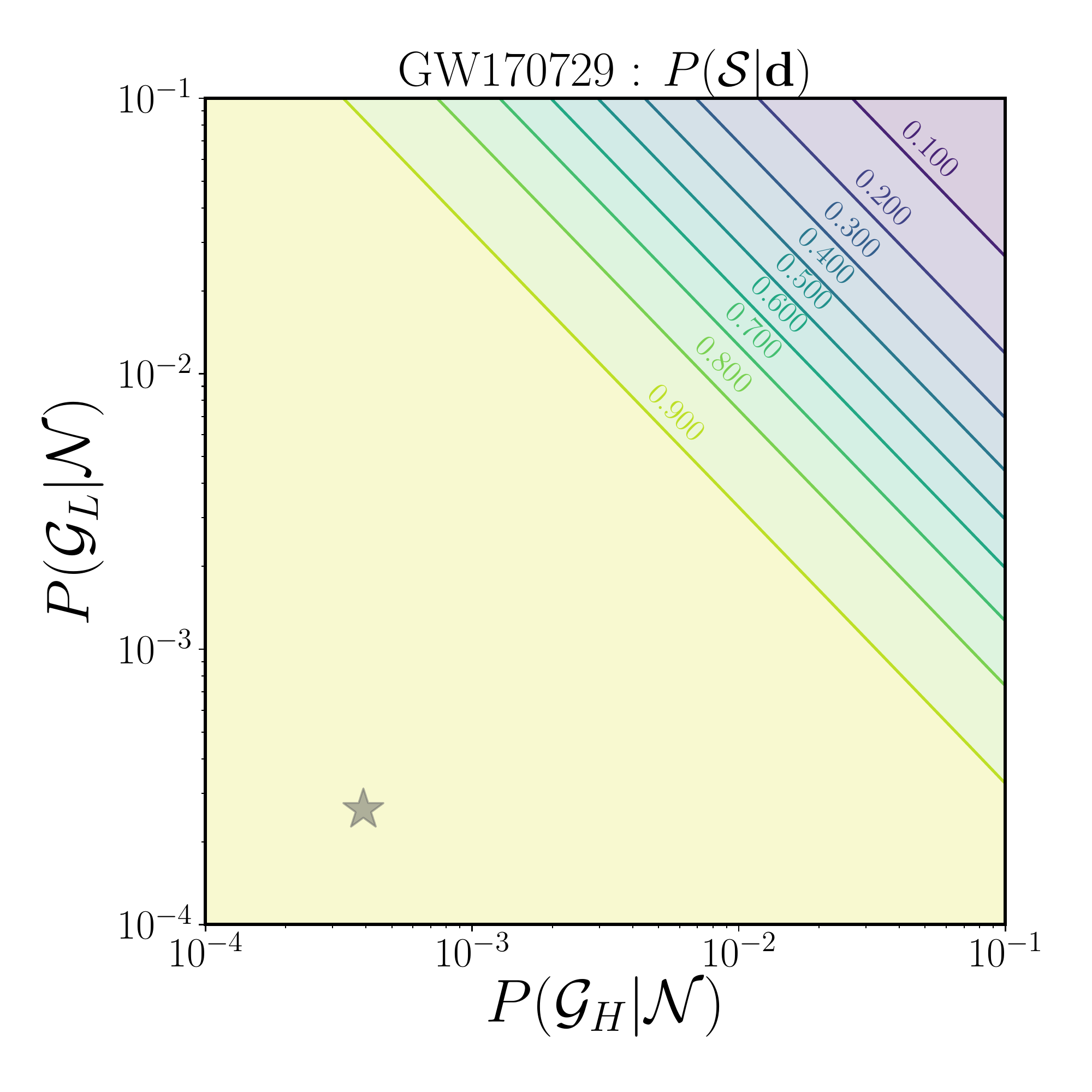}
    \includegraphics[width=0.8\columnwidth]{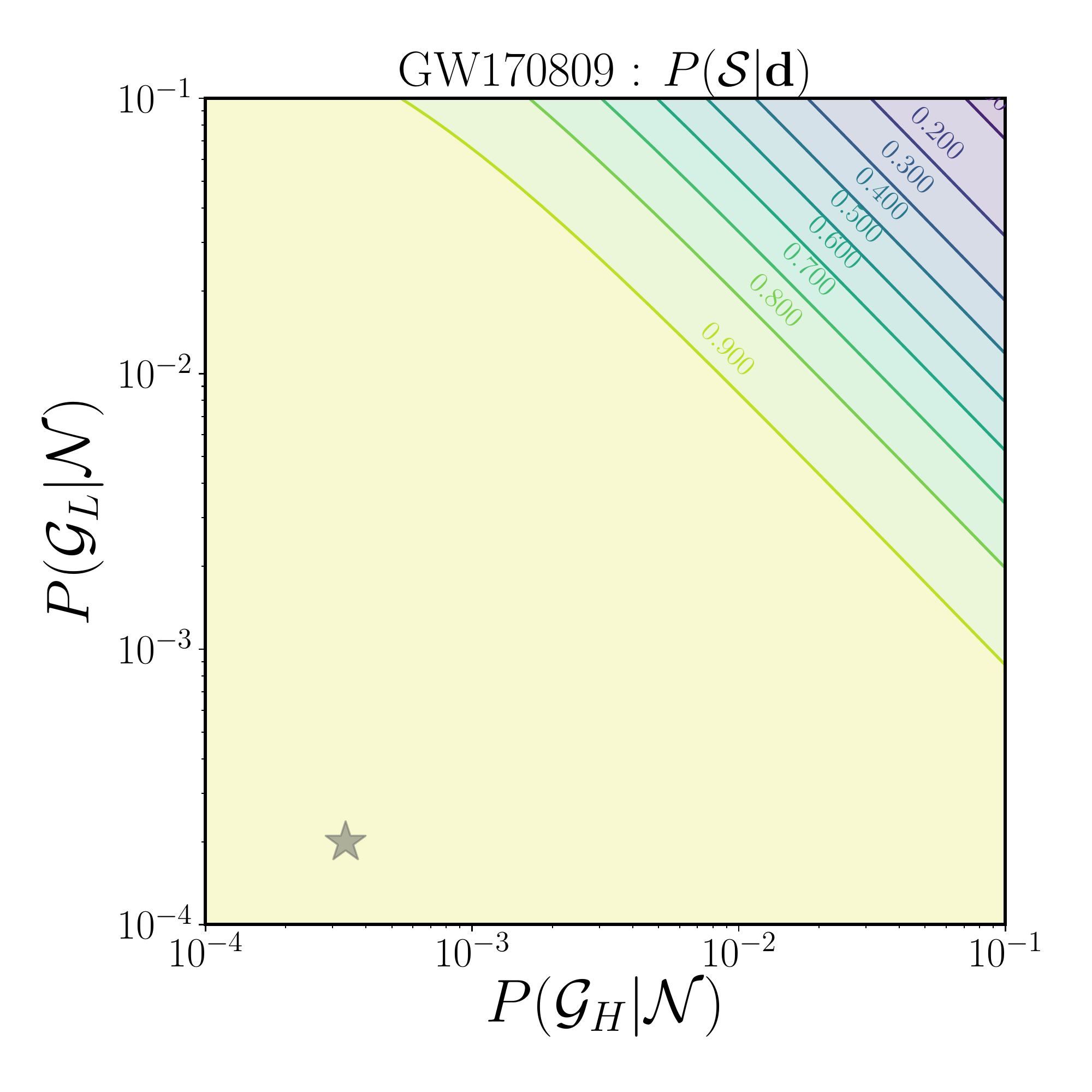}
        \caption{Posterior probability that the signal is of astrophysical origin $\pastro$ for select events as a function of the detector glitch probabilities $P(\mathcal{G}_H | \Hn)$ and $P(\mathcal{G}_L | \Hn)$ for the LIGO-Hanford and LIGO-Livingston instruments, respectively. The grey star denotes the glitch probability estimated from \omicron triggers, see Table~\ref{tab:PGk}. GW150914 is unambiguous, having an astrophysical probability $\approx 1$ irrespective of the glitch probability. Similarly, we find that GW170729 has $\pastro \approx 1$ for the glitch probabilities estimated from a $24$h window around the event. For the candidate events GW170121, GW170425 and GW170727, the probability that the signal is of astrophysical origin is more sensitive to prior probability on the glitch and signal hypotheses respectively. }
    \label{fig:pastro_contours}
\end{figure*}

In Fig.~\ref{fig:pastro_contours} we explore the dependence of $\pastro$ on the glitch prior for each of the detectors for selected events. The fiducial values of $P(\Hg_k | \dataall)$ estimated from the \omicron triggers are denoted with a grey star. 

\section{Posteriors for GW170121, GW170425 and GW170727}
\label{app:posteriors}
\begin{figure*}
    \centering
    \includegraphics[width=0.95\linewidth]{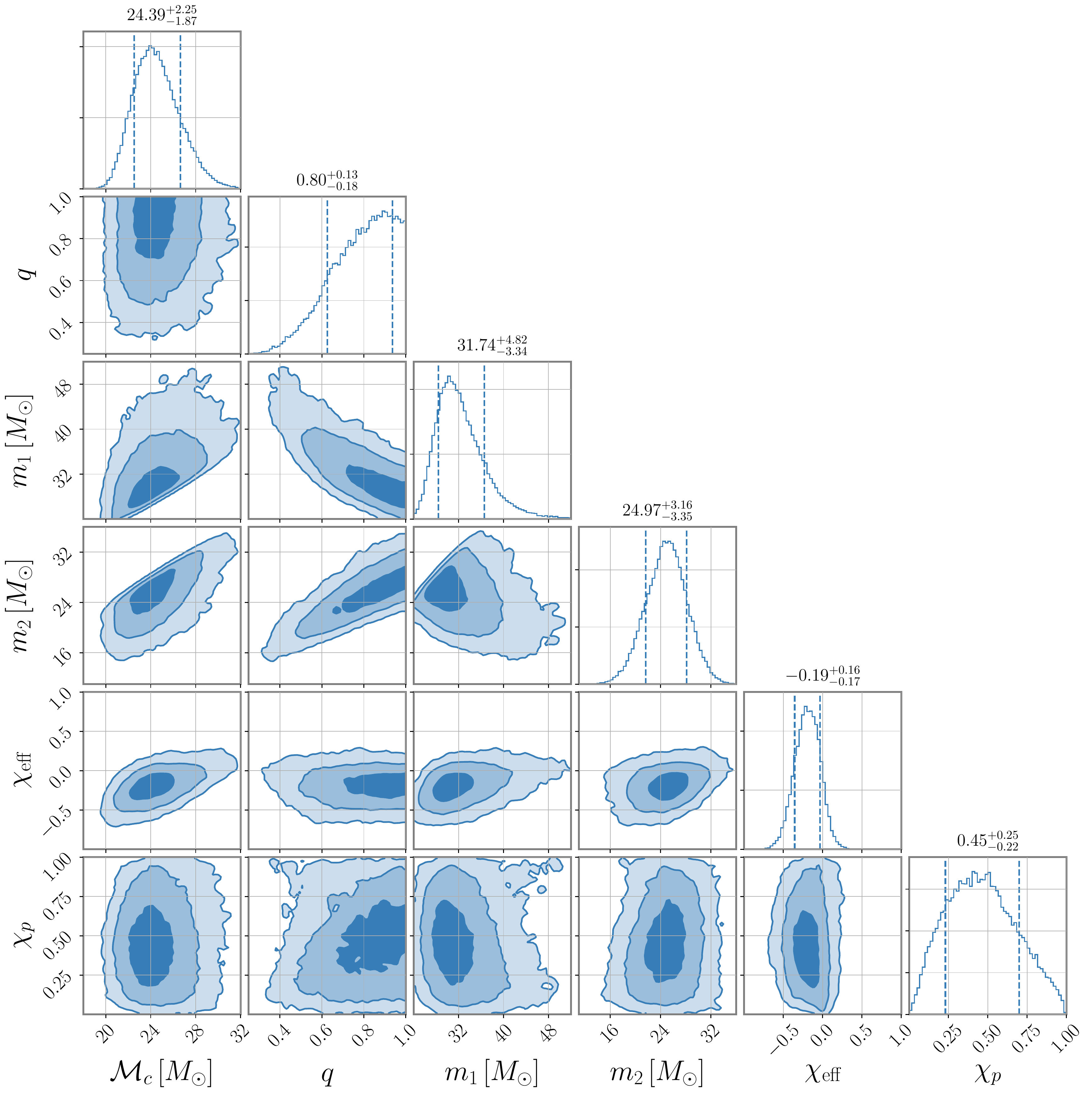}
        \caption{Posterior distributions for GW170121. We show source frame parameters: chirp mass $\mathcal{M}_c \equiv (m_1\,m_2)^{3/5}/(m_1 + m_2)^{1/5}$, mass ratio $q \equiv m_2/m_1 \le 1$, component masses $m_1$ and $m_2$, effective aligned spin $\chi_{\rm eff}$ and effective precessing spin $\chi_p$ \cite{Schmidt:2014iyl}, see e.g. Eq. (6) and (9), respectively, in~\cite{TheLIGOScientific:2016wfe}. Note that whilst GW170121 has negative effective spin $\chi_{\rm eff} \simeq -0.19$ \cite{Venumadhav:2019lyq,Huang:2020ysn}, the intrinsic parameters are broadly compatible with the population of BBHs observed in O1 and O2 \cite{LIGOScientific:2018mvr,LIGOScientific:2018jsj}.}
    \label{fig:corner_new_events_GW170121}
\end{figure*}

\begin{figure*}
    \centering
    \includegraphics[width=0.95\linewidth]{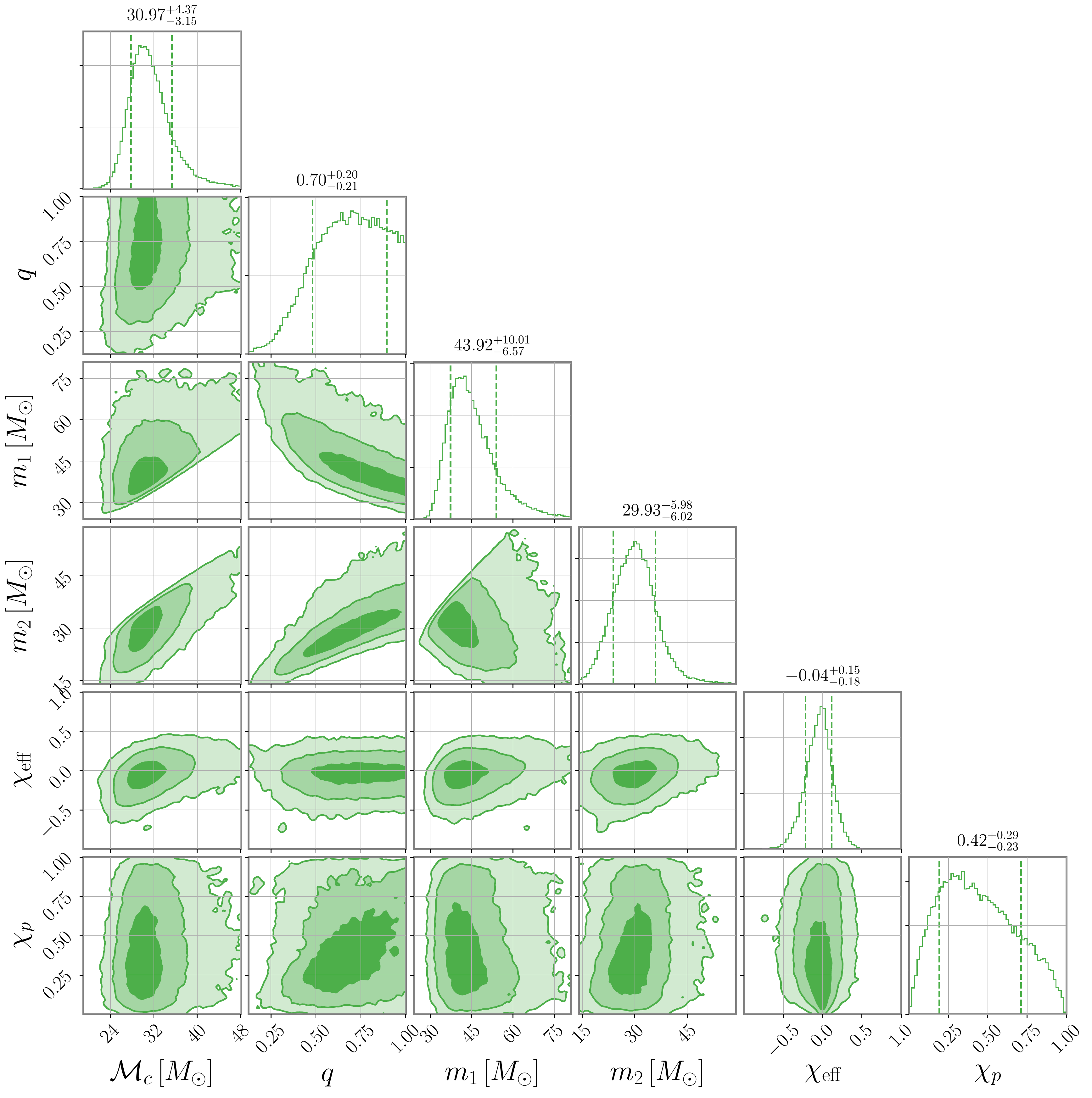} 
        \caption{Posterior distributions for GW170425 as per Fig.~\ref{fig:corner_new_events_GW170121}.}
    \label{fig:corner_new_events_GW170425}
\end{figure*}

\begin{figure*}
    \centering
    \includegraphics[width=0.95\linewidth]{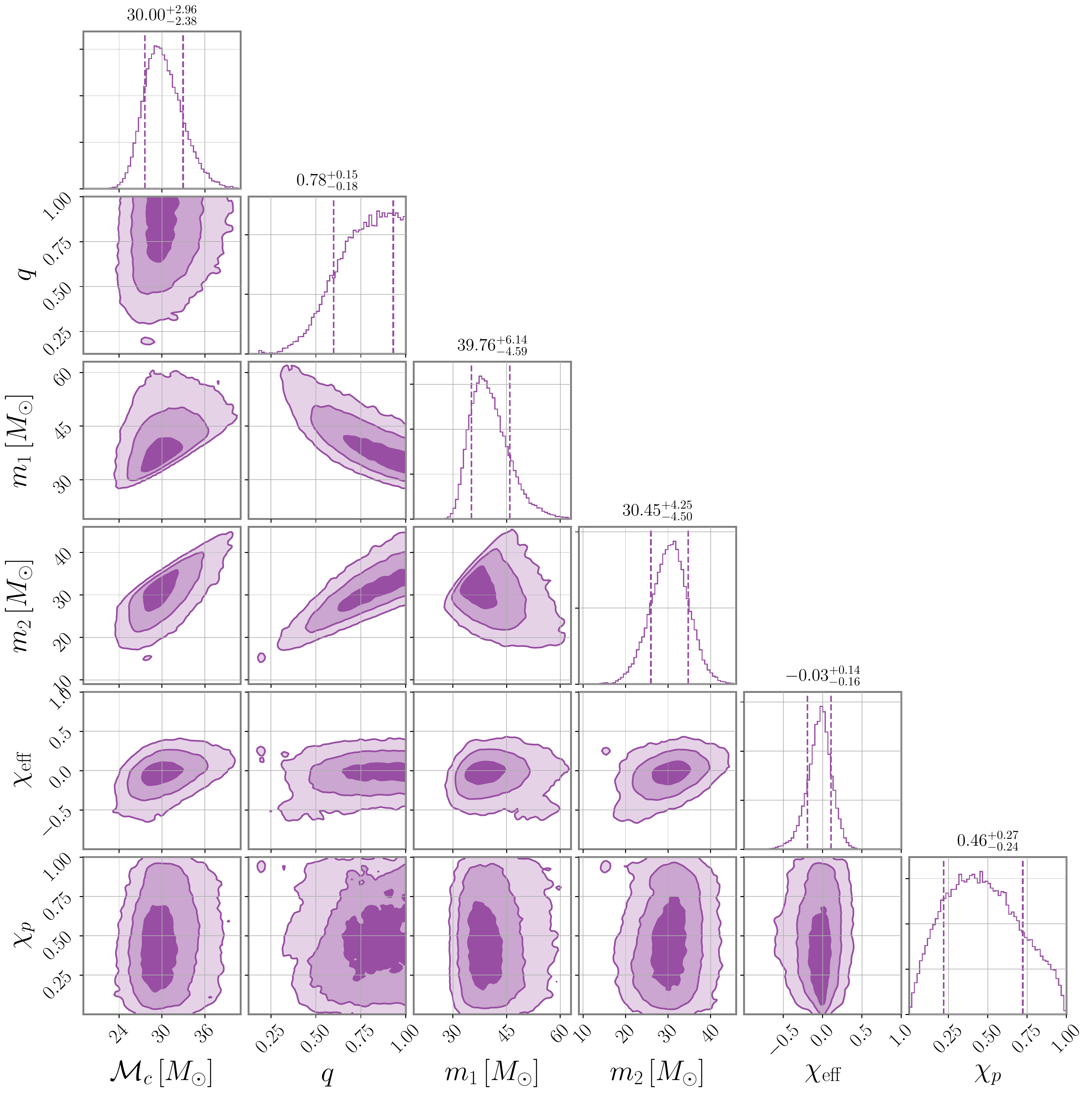}
        \caption{Posterior distributions for GW170727 as per Fig.~\ref{fig:corner_new_events_GW170121}.}
    \label{fig:corner_new_events_GW170727}
\end{figure*}

In Fig.~\ref{fig:corner_new_events_GW170121}, \ref{fig:corner_new_events_GW170425} and \ref{fig:corner_new_events_GW170727} we show the posterior distributions for a subset of key intrinsic parameters for the three candidate events with $\pastro > 0.5$: GW170121, GW170425 and GW170727. The posterior distributions show that the instrinsic parameters of the BBHs that generated these systems are broadly compatible with those that characterise the population of BBHs observed in O1 and O2 as reported in GWTC-1~\cite{LIGOScientific:2018mvr,LIGOScientific:2018jsj}, though we note that GW170121 has a negative effective aligned spin paameter $\chi_{\rm eff} \simeq -0.22^{+0.17}_{-0.17}$ \cite{Venumadhav:2019lyq,Huang:2020ysn}. 

\begin{figure*}[!htpb]
    \centering
    \includegraphics[width=0.99\textwidth]{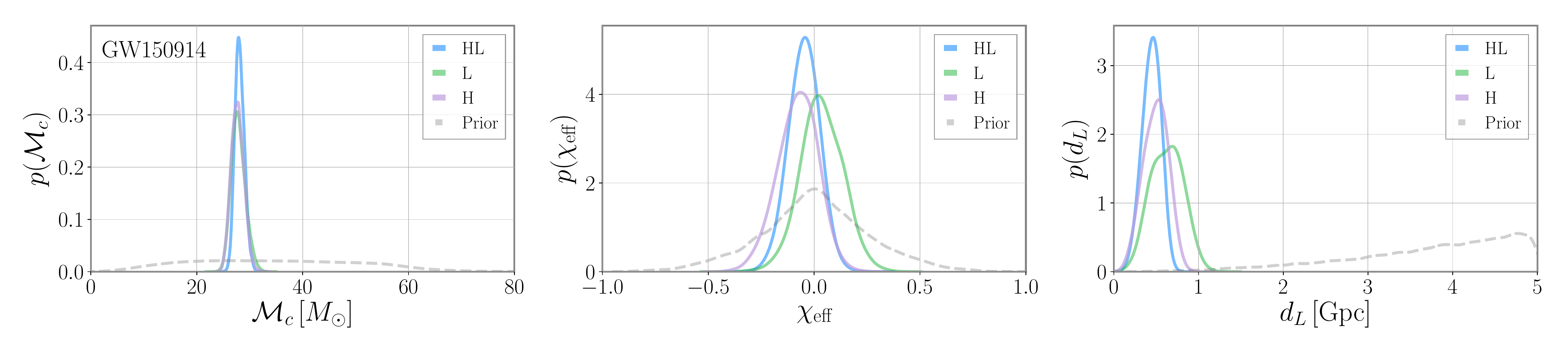}
    \includegraphics[width=0.99\textwidth]{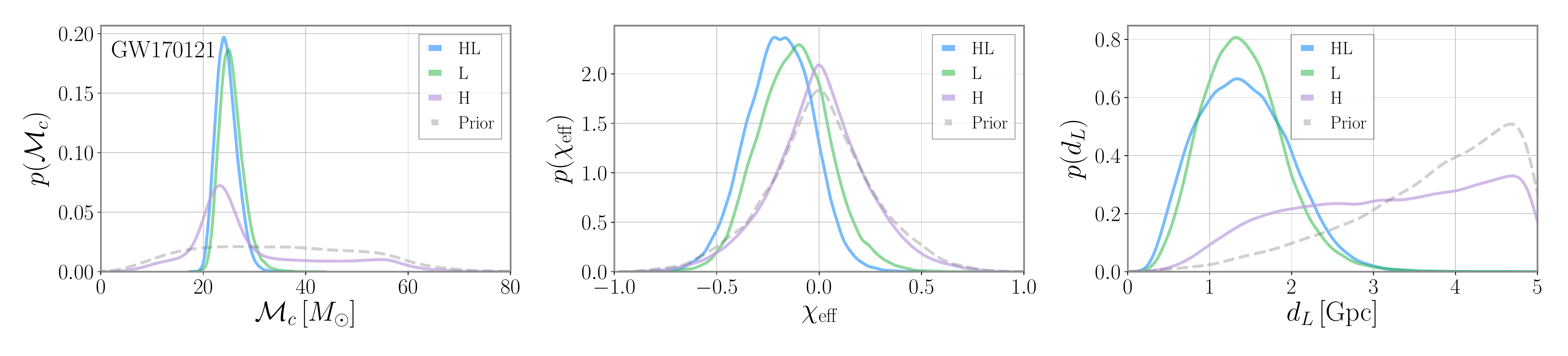}
    \includegraphics[width=0.99\textwidth]{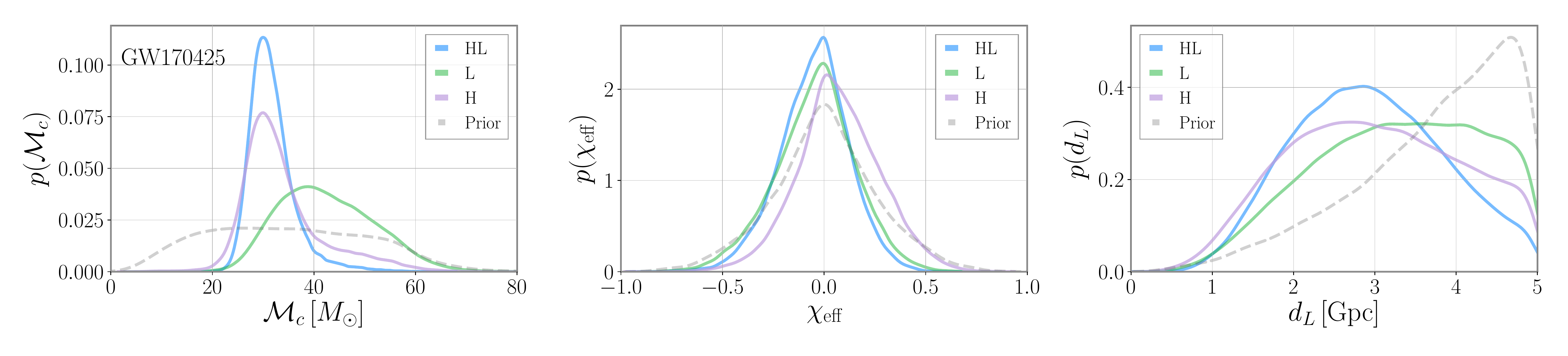}
    \includegraphics[width=0.99\textwidth]{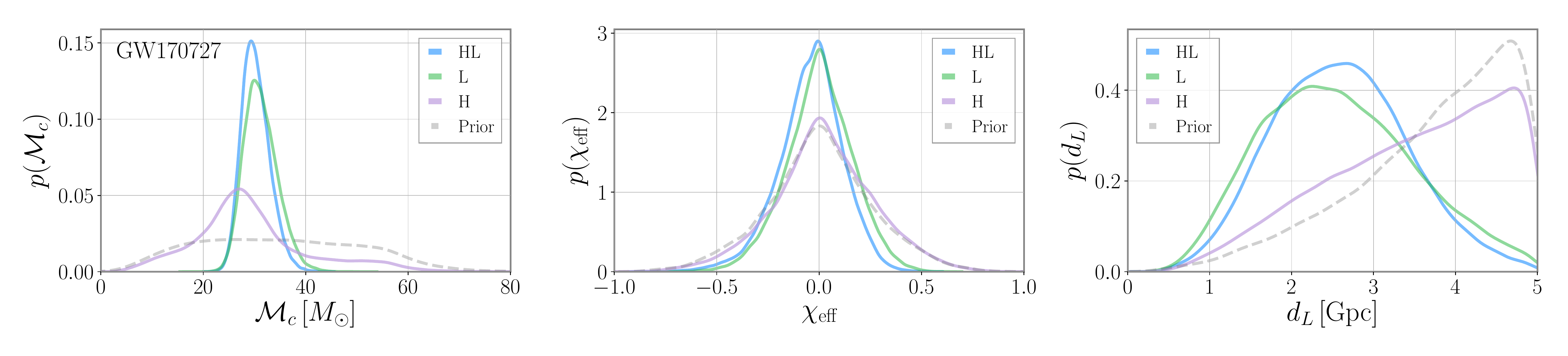}
    \includegraphics[width=0.99\textwidth]{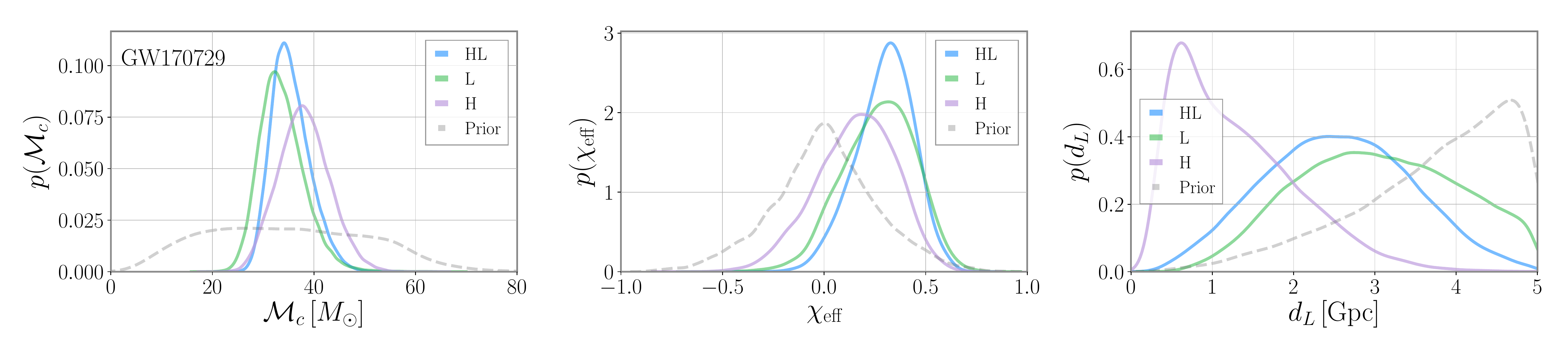}
    \caption{For coherent astrophysical signals, the posterior distributions for parameters that can be best measured from the data should be broadly consistent between the values inferred individually from each of the detectors and from the coherent network analysis. Here we show the posteriors for the chirp mass $\mathcal{M}_c$ and effective spin $\chi_{\rm eff}$ for the three candidate events GW170121, GW170425 and GW170727 as well as the GWTC-1 events GW150914 and GW170729. All events show evidence for posterior information being driven by both detectors. }
    \label{fig:posterior_consistency}
\end{figure*}

In Fig.~\ref{fig:posterior_consistency}, we perform a sanity check on the posterior distributions for the three events with $\pastro > 0.5$, GW170729 -- which was reported in GWTC-1 with different values of $p_\mathrm{astro}$ by different pipelines -- and GW150914, an unambiguous BBH. For a coherent astrophysical signal, the posterior distributions for parameters that can be well constrained from the data should be broadly consistent between the values inferred from each detector individually as well as the coherent network analysis. GW150914 and GW170728 both show significant information gain from Hanford and Livingston respectively. The posteriors for the new events of interest show that information is gained from each detector individually. 

%%%%%%%%%%%%%%%%%%%%%%%%%%%%%
\begin{table*}
\caption{The first three columns show the network signal-to-noise ratio and single detector signal-to-noise ratio (the indices ``L" and ``H" stand for LIGO-Livigston and LIGO-Hanford, respectively) for all the events considered in this analysis, see Tab.~\ref{tab:triggers}. The next three columns show the Bayes factors for the signal vs noise hypothesis for the detector network, Eq.~(\ref{eq:B_sn}), and the single detectors, Eq.~(\ref{eq:B_sn_k}), respectively. The last two columns report the glitch probabilities as estimated from \omicron triggers in a 24h stretch of data near each trigger. The glitch rates are determined by down-selecting \omicron triggers based on a single detector signal-to-noise ratio threshold of $\rho > 7$ and by restricting the frequencies to $20 < f_{\textsc{Om}} < 1024$Hz, see Eq.~(\ref{eq:PGk}) and discussion in Sec.~\ref{sec:priors}.}
\label{tab:PGk}
\centering
\begin{tabularx}{\linewidth}{SC@{} SC@{} SC@{} SC@{} SC@{} SC@{} SC@{} SC@{} SC@{} } 
\toprule
Event & $\rho^{\rm Net.}_{\rm MF}$ & $\rho^{H}_{\rm MF}$ & $\rho^{L}_{\rm MF}$ & $\log_{10} \mathcal{B}_{s/n}$ & $\log_{10} \mathcal{B}^{(H)}_{s/n}$ & $\log_{10} \mathcal{B}^{(L)}_{s/n}$ & $P(\Hg_H | \Hn)$  & $P(\Hg_L | \Hn)$  \\
\hline
\hline
\midrule
\rowcolor{austriawien!05}
GW150914 &             25.01 &             20.30 &             14.71 &                         121.1 &                               77.77 &                               36.45 &                       $5.35\times 10^{-4}$ &                       $1.23\times 10^{-4}$ \\
\rowcolor{austriawien!05}
GW151012 &              9.63 &              7.23 &              6.38 &                          8.25 &                                1.97 &                                0.77 &                       $8.73\times 10^{-4}$ &                       $3.61\times 10^{-4}$ \\
\rowcolor{austriawien!05}
GW151226 &             12.71 &             10.30 &              7.35 &                         18.45 &                                9.26 &                                0.23 &                       $9.42\times 10^{-4}$ &                       $1.74\times 10^{-4}$ \\
\rowcolor{austriawien!05}
GW170104 &             14.01 &              9.75 &             10.27 &                         30.19 &                               10.96 &                               12.85 &                       $6.16\times 10^{-4}$ &                       $4.42\times 10^{-4}$ \\
\rowcolor{austriawien!05}
GW170608 &             15.62 &             12.68 &              9.82 &                          34.7 &                               19.45 &                                7.46 &                       $5.00\times 10^{-4}$ &                       $3.82\times 10^{-4}$ \\
\rowcolor{austriawien!05}
GW170729 &             10.62 &              9.15 &              7.34 &                         15.52 &                                 7.4 &                                5.46 &                       $3.91\times 10^{-4}$ &                       $2.62\times 10^{-4}$ \\
\rowcolor{austriawien!05}
GW170809 &             12.82 &              7.77 &             11.01 &                         23.44 &                                3.47 &                               16.88 &                       $3.36\times 10^{-4}$ &                       $1.99\times 10^{-4}$ \\
\rowcolor{austriawien!05}
GW170814 &             16.66 &              9.27 &             14.08 &                         46.32 &                                8.96 &                               31.56 &                       $2.50\times 10^{-4}$ &                       $2.22\times 10^{-4}$ \\
\rowcolor{austriawien!05}
GW170818 &             11.34 &              3.70 &             10.58 &                         15.58 &                                0.05 &                               13.97 &                       $2.96\times 10^{-4}$ &                       $2.27\times 10^{-4}$ \\
\rowcolor{austriawien!05}
GW170823 &             12.04 &              7.19 &              9.93 &                         21.72 &                                3.97 &                               13.54 &                       $2.06\times 10^{-4}$ &                       $1.67\times 10^{-4}$ \\
\hline 
GW151011  &              7.46 &              4.05 &              6.18 &                           4.3 &                                0.28 &                                2.45 &                       $8.73\times 10^{-4}$ &                       $3.31\times 10^{-4}$ \\
GW151124  &              8.67 &              4.99 &              8.80 &                           5.1 &                               -0.28 &                                7.76 &                       $4.40\times 10^{-4}$ &                       $6.90\times 10^{-4}$ \\
GW151205  &              6.81 &              5.43 &              4.77 &                          4.67 &                                1.57 &                                1.04 &                       $6.81\times 10^{-4}$ &                       $7.55\times 10^{-4}$ \\
GW151216  &              8.16 &              5.41 &              6.10 &                          3.84 &                                0.36 &                                0.51 &                       $9.24\times 10^{-4}$ &                       $1.81\times 10^{-4}$ \\
GW151216A &              8.48 &              6.88 &              4.52 &                          0.18 &                               -0.08 &                               -0.09 &                       $9.24\times 10^{-4}$ &                       $1.71\times 10^{-4}$ \\
GW151217  &              7.99 &              5.59 &              3.55 &                          0.15 &                               -0.11 &                                0.04 &                       $9.24\times 10^{-4}$ &                       $1.71\times 10^{-4}$ \\
GW151222  &             10.22 &             10.50 &              3.50 &                          9.34 &                               10.86 &                                0.13 &                       $9.24\times 10^{-4}$ &                       $1.71\times 10^{-4}$ \\
GW170104A &              7.36 &              5.76 &              5.13 &                          1.37 &                                0.05 &                                0.34 &                       $6.16\times 10^{-4}$ &                       $4.42\times 10^{-4}$ \\
GW170106  &              9.62 &             10.05 &              2.90 &                          8.19 &                               11.15 &                               -0.09 &                       $1.58\times 10^{-3}$ &                       $1.05\times 10^{-3}$ \\
\rowcolor{dodgerblue!10}
GW170121  &             10.55 &              5.43 &              9.04 &                         12.26 &                                 0.4 &                                8.49 &                       $2.78\times 10^{-4}$ &                       $3.42\times 10^{-3}$ \\
GW170123  &              6.20 &              4.29 &              5.61 &                          2.34 &                                0.05 &                                0.58 &                       $1.85\times 10^{-4}$ &                       $1.08\times 10^{-3}$ \\
GW170201  &              8.25 &              6.44 &              6.12 &                          3.64 &                                0.36 &                                0.45 &                       $2.64\times 10^{-4}$ &                       $3.19\times 10^{-4}$ \\
GW170202  &              8.50 &              5.26 &              7.29 &                          5.33 &                                0.31 &                                1.48 &                       $3.82\times 10^{-4}$ &                       $3.10\times 10^{-4}$ \\
GW170220  &              6.59 &              4.68 &              4.81 &                          2.18 &                                0.18 &                                0.94 &                       $2.57\times 10^{-4}$ &                       $2.71\times 10^{-4}$ \\
GW170304  &              8.50 &              5.75 &              7.50 &                          7.53 &                                0.46 &                                5.83 &                       $2.20\times 10^{-4}$ &                       $1.04\times 10^{-3}$ \\
GW170402  &              9.00 &              3.43 &              9.29 &                          6.16 &                                0.01 &                                6.67 &                       $2.92\times 10^{-4}$ &                       $7.04\times 10^{-4}$ \\
GW170403  &              8.20 &              6.27 &              6.47 &                           5.8 &                                1.56 &                                1.79 &                       $4.00\times 10^{-4}$ &                       $9.21\times 10^{-4}$ \\
\rowcolor{dodgerblue!10}
GW170425  &              7.99 &              5.75 &              5.98 &                          6.74 &                                1.69 &                                2.61 &                       $5.99\times 10^{-4}$ &                       $4.56\times 10^{-4}$ \\
GW170620  &              8.36 &              6.05 &              6.11 &                          3.38 &                                0.11 &                                0.33 &                       $8.77\times 10^{-4}$ &                       $3.06\times 10^{-4}$ \\
GW170629  &              7.56 &              6.40 &              3.41 &                         -0.06 &                               -0.12 &                                -0.3 &                       $7.62\times 10^{-4}$ &                       $3.91\times 10^{-4}$ \\
GW170721  &              8.55 &              7.13 &              5.88 &                          5.66 &                                3.08 &                                0.23 &                       $5.51\times 10^{-4}$ &                       $2.87\times 10^{-4}$ \\
GW170724  &             10.76 &             11.84 &              6.27 &                         13.46 &                               17.14 &                                -0.2 &                       $7.87\times 10^{-4}$ &                       $2.69\times 10^{-4}$ \\
\rowcolor{dodgerblue!10}
GW170727  &              9.90 &              4.97 &              7.66 &                          9.15 &                                0.16 &                                6.27 &                       $4.54\times 10^{-4}$ &                       $2.52\times 10^{-4}$ \\
GW170729A &             11.55 &             11.17 &              4.35 &                          10.5 &                               12.73 &                                 0.7 &                       $3.31\times 10^{-4}$ &                       $2.89\times 10^{-4}$ \\
GW170801  &              8.14 &              3.14 &              8.05 &                          2.69 &                                0.05 &                                4.17 &                       $2.45\times 10^{-4}$ &                       $2.22\times 10^{-4}$ \\
GW170817A &             10.65 &              3.82 &             10.42 &                         16.62 &                                0.11 &                               15.88 &                       $2.55\times 10^{-4}$ &                       $1.62\times 10^{-4}$ \\
\hline
\rowcolor{austriawien!05}
GW190412  &             18.47 &              9.77 &             15.80 &                         59.27 &                               10.59 &                               42.54 &                       $1.00\times 10^{-4}$ &                       $1.00\times 10^{-4}$ \\
\bottomrule
\end{tabularx}
\end{table*}

%%%%%%%%%%%%%%%%%%%%%%%%%%%%%%%%
\newpage

%%%%%%%%%%%%%%%%%%%%%%
\newpage

%%%%%%%%%%%%%%%%%%%%%%
\newpage

%%%%%%%%%%%%%%%%%%%%%%
\newpage

\newpage

\clearpage
%%%%%%%%%%%%%%% BIBLIOGRAPHY
\bibliographystyle{apsrev4-1}
\bibliography{refs}

\end{document}